\documentclass[iop]{emulateapj}

\newcommand{\nuvr}{NUV$-r$\,\,}

\newcommand{\sphy}{$\Sigma_\mathrm{1}$}

\newcommand{\vsc}{$\sigma_{1}$}

\usepackage{natbib}
\usepackage{apjfonts}
\usepackage[colorlinks=true,citecolor=blue]{hyperref}
\usepackage{graphicx}
\usepackage[usenames,dvipsnames]{color}
\definecolor{darkgreen}{RGB}{0,100,0}

\shorttitle{SF QUENCHING AND INNER STELLAR MASS DENSITY}
\shortauthors{FANG ET AL.}

\begin{document}

\title{A Link Between Star Formation Quenching and Inner Stellar Mass Density in SDSS Central Galaxies}

\author{Jerome J.~Fang\altaffilmark{1}, S.~M.~Faber\altaffilmark{1}, David~C.~Koo\altaffilmark{1}, \& Avishai Dekel\altaffilmark{2}}

\altaffiltext{1}{UCO/Lick Observatory, Department of Astronomy and Astrophysics, University of California, Santa Cruz, CA 95064, USA; {jjfang@ucolick.org}}
\altaffiltext{2}{Racah Institute of Physics, The Hebrew University, Jerusalem 91904, Israel}

\submitted{ApJ Accepted, 2013 August 23}

\begin{abstract}

We study the correlation between galaxy structure and the quenching of star formation using a sample of SDSS central galaxies with stellar masses $9.75<\log M_*/M_\odot<11.25$ and redshifts $z<0.075$. GALEX UV data are used to cleanly divide the sample into star-forming and quenched galaxies, and to identify galaxies in transition (the green valley). Despite a stark difference in visual appearance between blue and red galaxies, their average radial \emph{stellar mass density} profiles are remarkably similar (especially in the outer regions) at fixed mass. The inner stellar mass surface density within a radius of 1 kpc, \sphy, is used to quantify the growth of the bulge as galaxies evolve. When galaxies are divided into narrow mass bins, their distribution in the color-\sphy\ plane at fixed mass forms plausible evolutionary tracks. \sphy\ seems to grow as galaxies evolve through the blue cloud, and once it crosses a threshold value, galaxies are seen to quench at fixed \sphy. The \sphy\ threshold for quenching grows with stellar mass, $\Sigma_1\propto M_*^{0.64}$. However, the existence of some star-forming galaxies above the threshold \sphy\ implies that a dense bulge is necessary but not sufficient to quench a galaxy fully. This would be consistent with a two-step quenching process in which gas \emph{within} a galaxy is removed or stabilized against star formation by bulge-driven processes (such as a starburst, AGN feedback, or morphological quenching), whereas \emph{external} gas accretion is suppressed by separate halo-driven processes (such as halo gas shock heating). Quenching thus depends on an interplay between the inner structure of a galaxy and its surrounding dark matter halo, and lack of perfect synchrony between the two could produce the observed scatter in color vs.~\sphy. Finally, we find a tight correlation between \sphy\ and the velocity dispersion scaled to a 1-kpc aperture, \vsc, for all galaxies, $\Sigma_1\propto\sigma_1^{2.0}$. This relation is used to predict a new black hole mass scaling relation, $M_\mathrm{BH}\propto\Sigma_1^{2.0}$, that can be tested with existing observations of nearby galaxies.

\end{abstract}


\section{INTRODUCTION} \label{introduction}

Our knowledge about galaxy formation and evolution has blossomed in recent years thanks to ever-larger observational surveys and increasingly sophisticated theoretical models. However, a full understanding of the processes by which star formation (SF) shuts down in galaxies and their connection to galaxy structure remains elusive. Observationally, galaxies exhibit a bimodal distribution in color that persists even at high redshifts \citep[e.g.,][]{strateva01,kauffmann03,brammer09}. The bimodality has been interpreted as an evolutionary path: blue and star-forming galaxies undergo one or more processes that ultimately shut down (or ``quench'') SF and cause them to migrate toward redder colors \citep{bell04,faber07,martin07}. Much work has been done to elucidate possible quenching processes, primarily from a theoretical perspective. Our focus in this work is on quenching in \emph{central} galaxies, so we do not discuss processes that can quench SF in satellite galaxies.

The ultimate source of fuel for SF is (molecular) gas. In order to quench SF, the process(es) must remove such gas, heat it/prevent it from cooling, or stabilize it against gravitational collapse. Powerful AGN feedback (i.e., ``quasar mode'' feedback) is widely invoked as a means to expel gas from a galaxy \citep[e.g.,][]{dimatteo05,hopkins06}. Such feedback is predicted to be the result of a major gas-rich merger that initially fuels a powerful starburst as well as later activates the AGN, which subsequently blows out any remaining gas, thereby quenching SF. After the quasar phase, low-level AGN feedback continues to heat the surrounding gas and prevents it from cooling and settling (back) into the galaxy \citep[i.e., ``radio mode'' feedback;][]{croton06}. The combination of these two AGN-driven processes effectively stifles any future SF.

An alternative method to suppress SF is to make any gas stable against gravitational collapse. In such a process, gas does not have to be expelled from the galaxy itself. Recently, \citet{martig09} proposed that the buildup of a central bulge is sufficient to stabilize the surrounding gas disk from collapsing and forming stars. Stability is achieved because the deeper gravitational potential well of the bulge increases the shear in the disk, making it difficult for the gas to form bound clumps (i.e., its Toomre $Q>1$). Such ``morphological quenching'' can persist over several Gyr, and it is effective even if gas is continually accreted from external sources.

The quenching processes discussed so far operate \emph{within} a galaxy. Of course, galaxies reside within dark matter halos, and it is not surprising that the halo itself may play a critical role in quenching SF. Simulations have shown that as halos reach a critical mass of $\sim10^{12} M_\odot$, infalling gas in the halo becomes shock-heated to the virial temperature and is no longer able to cool efficiently \citep[e.g.,][]{birnboim03,keres05,dekel06}. The formation of a virial shock thus prevents gas from accreting onto the galaxy, robbing it of fuel for additional SF.  Such ``halo quenching'' is not predicted to be a sharp, well-defined transition; theoretical predictions and observational evidence both indicate a large scatter in the critical halo mass of at least $\sim1$~dex \citep{keres05,dekel06,woo13}. Note that halo quenching does not require the presence of an AGN in order to operate.  

A complementary approach to understanding quenching via gas physics has been to empirically examine how galaxy structure changes as galaxies evolve from blue to red. This approach is motivated by observed correlations between various structural parameters (e.g., mass concentration, velocity dispersion) and some measure of SF activity (e.g., color, specific SF rate). Presumably, the change in such a structural parameter as a galaxy evolves would be indicative of the relevant quenching process(es) at work. 

Studies of SDSS galaxies have demonstrated the existence of critical or threshold values of structural parameters that define a transition in the recent SF histories of galaxies. For example, below a critical stellar mass, $M_*\sim3\times10^{10}\,M_\odot$, galaxies typically are young and actively forming stars, while above this value, galaxies are predominantly old and not star-forming \citep{kauffmann03a}. A similar threshold in effective stellar surface mass density $(\propto M_*/R_\mathrm{eff}^2)$ has also been seen: above a value of $\sim3\times10^{8}\,M_\odot\,\mathrm{kpc}^{-2}$, galaxies are generally old and quiescent, while the opposite is true below this threshold \citep{kauffmann03a,kauffmann06}. Moreover, \citet{kauffmann03a} showed that surface density, rather than stellar mass, is more strongly correlated with $D_n(4000)$, a proxy for galaxy age. This latter result is an indication that the \emph{distribution} of mass within a galaxy may be the key tracer (or driver?)~of quenching in galaxies. Indeed, a similar threshold has been found when examining the Sersic index \citep{sersic68}, which quantifies the mass (more precisely, the light) distribution \citep{driver06,schiminovich07,bell08}. In particular, galaxies above a Sersic value of $n\approx2.5$ tend to be quiescent.\footnote{However, as discussed in Section \ref{profiles}, the difference in Sersic index between blue and red galaxies is primarily due to the presence or lack of a bright, star-forming disk; it does not necessarily signal any large difference in the \emph{mass} distribution.} The existence of structural thresholds in galaxies has also been seen at higher redshift \citep{franx08,bell12,cheung12}.

The results discussed above focus on the \emph{general} galaxy population. For completeness, we mention here the existence of two classes of galaxies that have atypical quenching characteristics: post-starburst galaxies and low surface brightness galaxies. Post-starburst (or E+A) galaxies \citep{dressler83,quintero04,goto05} are rapidly quenched objects that likely underwent a recent merger-induced starburst \citep[e.g.,][]{hopkins06,snyder11}. These objects are rare in the local universe \citep[a few percent,][]{wong12,mendel13} and do not contribute much to the ``quenching budget'' at late times. However, they may represent the dominant quenching channel in the early universe \citep[e.g.,][]{whitaker12,barro13}. Understanding the connection between their structure and SF quenching is the subject of a forthcoming SDSS study (H.~Yesuf, in preparation). 

Low surface brightness galaxies are characterized by a central surface brightness fainter than $\approx22$ mag arcsec$^{-2}$ in $B$ \citep[e.g.,][]{impey97}. Detecting these objects has been greatly enhanced thanks to surveys like SDSS. With larger samples, it has been found that these galaxies span a wide range in e.g., colors, SF rates, masses, gas content, and sizes \citep{galaz02,galaz11,zhong08,zhong10}. This diversity makes it unclear if SF in these objects plods along at a very low rate \citep{wyder09}, or if they experience bursty episodes followed by periods of quiescence \citep{boissier08,zhong10}. Understanding how SF is quenched in low surface brightness galaxies remains an important issue, though in this work we do not discuss them in detail since they are a tiny fraction of our sample (see Section \ref{data}). 


Despite the existence of the structural thresholds discussed above, several challenges make it difficult to conclude that the structural configuration of a galaxy is sufficient to predict its SF history. First, the existence of thresholds in structural parameters does not immediately imply that \emph{all} galaxies quench once they reach such a threshold. This is borne out in the data as a non-negligible fraction of ``outlier'' galaxies that are above a putative threshold value, e.g., Sersic index $n>2.5$, yet are still blue and star-forming \citep{schiminovich07,bell08,bell12,cheung12}. In other words, being above the threshold is a necessary, but not sufficient, condition in order to be quenched. This suggests that quenching SF may be a multi-step process in which having the necessary galaxy structure is merely one component.

Another obstacle in linking structure with quenching is determining whether such thresholds are in fact fixed triggers (i.e., mass-independent and/or redshift-independent) over which galaxies evolve, or whether the thresholds evolve and ``sweep up'' galaxies (whose structural parameters were established early on and are now unchanging). Given that galaxies inevitably change their structure as they evolve, it is difficult to disentangle these two interpretations of thresholds.     

Yet another question is whether thresholds are causative or merely correlative, with both parameters descending from a separate, third cause. In this paper, we are conservative and infer only that a correlation exists and that causation has not yet been proved. Thus, in showing a strong correlation between quenching and inner mass surface density (Section \ref{sd1_nuvr}), we do not conclude that high mass density definitively causes quenching, but rather only that it tends to predict it.

A final challenge in connecting structure with quenching is deciding whether one structural parameter is superior to others in its ability to predict a galaxy's SF history and how that parameter is related to the actual physics of quenching. It is not expected that \emph{one} parameter can fully encapsulate all the relevant physical processes; nevertheless, one might gain insight into the \emph{dominant} mechanism at work by determining the most important structural parameter. It has been argued that Sersic index is the parameter that best discriminates between star-forming and quiescent galaxies \citep[][]{bell12}. However, \citet{wake12} conclude that velocity dispersion (measured within one-eighth of an effective radius) shows an even better correlation with galaxy color in SDSS galaxies. Regardless of the exact details, both Sersic index and velocity dispersion reflect the \emph{inner} mass distribution, which may therefore play a critical role in quenching SF. For example, both AGN feedback and morphological quenching rely on the presence of a sufficiently massive bulge in order to quench SF. A link to velocity dispersion might even exist in the halo quenching picture if it is indeed a better indicator of halo mass than stellar mass is \citep[][but see \citet{li13}]{wake12a}. But in this case the galactic structural parameter would signify the role of the halo rather than itself being the causative agent.

The growing appreciation that quenching correlates closely with galaxy structure has spurred further work recently. \citet{cheung12}, using a sample of galaxies at $z\sim0.8$ drawn from the AEGIS survey \citep{davis07}, investigated the utility of various structural parameters at predicting galaxy color. A novel feature was their use of the stellar surface mass density within a radius of 1 kpc, denoted \sphy\ hereafter, as a parameter to probe the inner regions of a galaxy. They found that \sphy\ shows an even tighter correlation with color than Sersic index, suggesting that it is the \emph{inner mass distribution} that is a critical indicator of quenching.

While the result of \citet{cheung12} is tantalizing, it remains to be verified with the larger samples available at lower redshifts. In this work, we harness the rich datasets available from the SDSS and GALEX surveys to study the correlation of \sphy\ with color found by \citet{cheung12} in more detail. Specifically, we examine trends in narrow mass slices as a way to uncover possible systematic trends with mass. A key feature of our work is the use of GALEX UV data rather than rest-frame optical $U-B$ color in order to better resolve the transition region between star-forming and quiescent galaxies (i.e., the green valley). This allows us to separate galaxies that have recently quenched from redder galaxies that presumably quenched at earlier epochs (and may have built up mass post-quenching via dry mergers). In addition, we take advantage of high-quality aperture photometry from SDSS that enables measurements of mass profiles for a large sample of galaxies as a function of mass and color.  

The paper is organized as follows. Our sources of data, sample selection, and method of computing mass profiles are discussed in Section \ref{data}. A novel and intriguing mass-dependent relation between inner surface density and \nuvr color is presented in Section \ref{sd1_nuvr}. Section \ref{profiles} presents average surface brightness and mass profiles and highlights our finding that mass buildup in the central regions is a key correlate of quenching in galaxies. We also compare the use of light-weighted vs.~mass-weighted quantities in studying galaxy structure. Section \ref{comparison} presents a comparison between inner mass density and velocity dispersion as tracers of quenching. Finally, our discussion and conclusions follow in Sections \ref{discussion} and \ref{conclusions}. All magnitudes in this paper are on the AB system \citep{oke74}. We assume a concordance $\Lambda$CDM cosmology with $\Omega_m=0.3,\Omega_\Lambda=0.7,$ and $H_0=70\mathrm{\,km\,s^{-1}\,Mpc^{-1}}$.

\section{DATA AND SAMPLE SELECTION} \label{data}

\subsection{Photometry and Structural Parameters} \label{parameters}

For this study, six-band integrated photometry (NUV from GALEX, $ugriz$ from SDSS) was obtained from the cross-matched GALEX GR6/SDSS DR7 catalog available through the GALEX CASJobs interface\footnote{\texttt{http://galex.stsci.edu/casjobs/}}. For GALEX magnitudes, \texttt{mag\_auto} values are used; for SDSS, model magnitudes are used. The photometry was corrected for Galactic extinction and $k$-corrected to $z=0$ using version 4.2 of the \texttt{kcorrect} code package described in \citet{blanton07}. 

In addition to integrated photometry, the SDSS pipeline measures surface brightness profiles in all five ($ugriz$) bands in a series of circular annuli of fixed angular size. For this work, the apertures chosen range in radius from 0\farcs23 to 11\farcs42. The annular photometry was corrected for Galactic extinction and $k$-corrected by computing independent $k$-corrections in each annulus. The $k$-corrected surface brightness profiles were then smoothly interpolated following the method outlined in \citet{stoughton02}. Briefly, a spline was fit to the \emph{cumulative} light profile (in order to conserve flux), which was then differentiated to obtain the final surface brightness profile interpolated over a fixed grid of angular sizes in 0\farcs1 intervals.

Spectroscopic redshifts, fiber velocity dispersions, and total stellar masses were obtained from the MPA/JHU DR7 value-added catalog\footnote{\texttt{http://www.mpa-garching.mpg.de/SDSS/DR7/}}. The stellar masses were computed by fitting the integrated SDSS photometry with stellar population models [similar in spirit to the method used in \citet{salim07}]. We additionally obtain the group membership (central or satellite) of the galaxies using the catalog of \citet{yang12}. 

\subsection{Sample Selection} \label{sample}

Our sample was selected from the GALEX GR6/SDSS DR7 cross-matched catalog. The sample was selected (1) to have redshift $0.005<z<0.12$, (2) to be within the central 0\fdg55 of the GALEX field of view, and (3) to be detected in the NUV bandpass of the GALEX Medium Imaging Survey. In addition, to exclude dusty, star-forming galaxies whose dust-reddened colors would move them into the green valley, only galaxies with axis ratio $b/a>0.6$ were kept. Finally, since we are interested in the quenching of central galaxies, not satellites, we selected only objects listed as the most massive group member in the SDSS DR7 group catalog of \citet{yang12}.

\begin{figure}
	\epsscale{1.25}
	\plotone{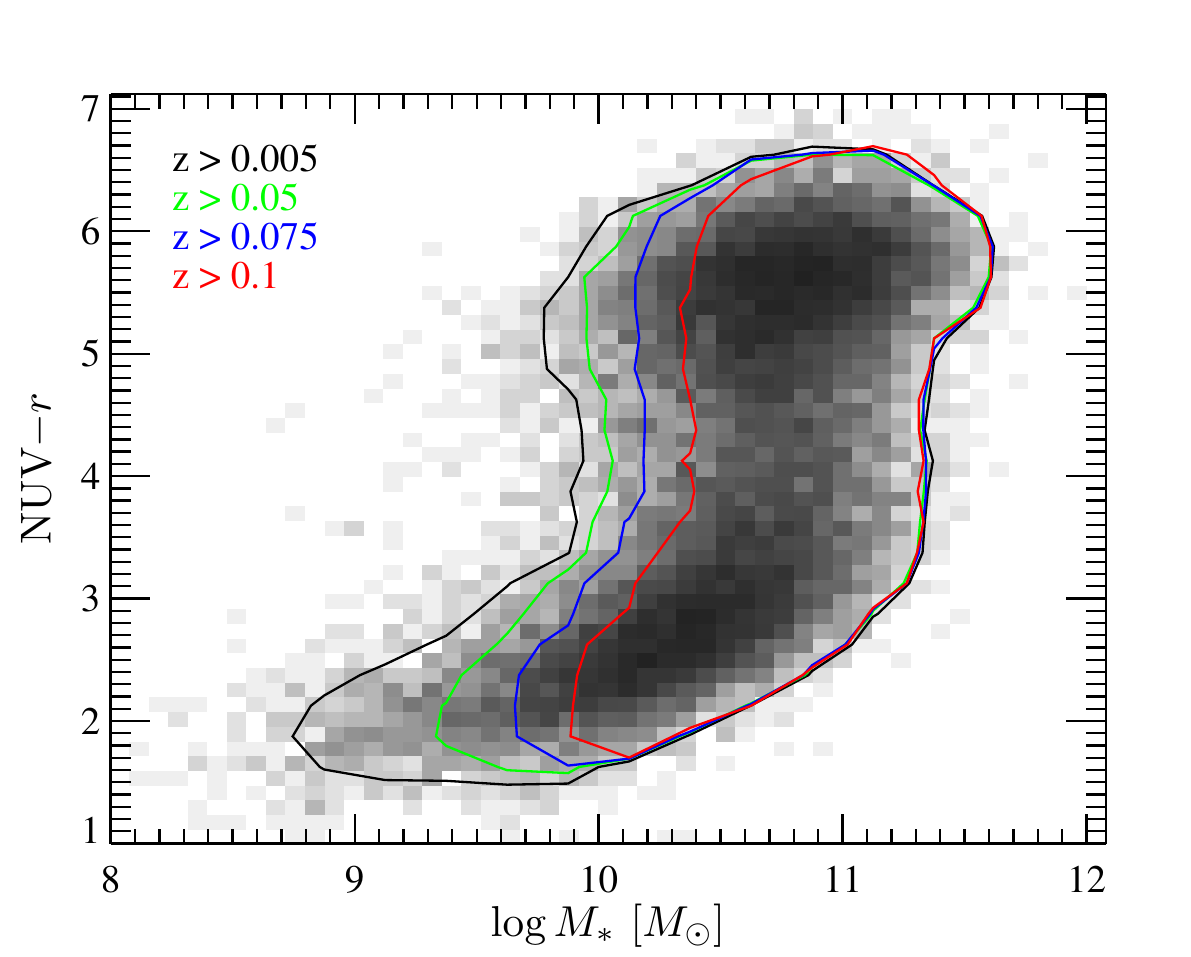}
	
	\caption{\nuvr color vs.~stellar mass for SDSS galaxies with GALEX NUV detections. The 2D histogram (grayscale) shows the distribution of central, face-on ($b/a>0.6$) galaxies with redshift $0.005<z<0.12$, clearly showing the well-defined blue and red sequences, with the green valley located in between at intermediate colors. The contours enclose those galaxies located above the redshift limits listed at upper left, indicating the approximate completeness limit at those redshifts. In this work, our final sample is restricted to $z<0.075$.} 

	\label{cmd_plot}
\end{figure}

Figure \ref{cmd_plot} presents the \nuvr color-stellar mass diagram for our sample. The sensitivity of the UV to even low-level SF is clearly demonstrated: galaxies cleanly separate into two well-delineated populations (the blue and red sequences\footnote{In this paper, the terms ``blue sequence'' and ``blue cloud'' are used interchangeably. The former is a more accurate description of the distribution of star-forming galaxies in a UV-optical color-mass diagram.}), with a clear transition region in between (the green valley). Included in Figure \ref{cmd_plot} are contours indicating the completeness of the sample above a given redshift (indicated in the figure). Each contour encloses 95\% of the total points in each redshift range. The loss of lower-mass galaxies with increasing redshift is due to the SDSS spectroscopic limit ($r=17.77$) and the GALEX NUV magnitude limit (NUV = 23). 

Throughout this paper we divide the sample into mass slices 0.25 dex wide in this color-stellar mass diagram. We emphasize that this is possible thanks to the large number of galaxies available in SDSS. Dividing by mass can uncover trends that are washed out when considering a wide range of stellar masses all together. An implicit assumption made later on is that galaxies evolve and redden without significantly increasing their stellar mass. Indeed, existing quenching scenarios predict either very little stellar mass growth or at most a doubling of stellar mass (via major mergers) between the blue and red sequences (see Section \ref{sd1_nuvr} for further discussion). In addition, one benefit of using \nuvr color rather than an optical color is that the former is much more sensitive to low-level SF and hence can be used to identify galaxies that have \emph{recently} quenched. The clear green valley seen in Figure \ref{cmd_plot} is dominated by those galaxies that are currently transitioning from star-forming to quiescent.  

\begin{figure}
	\epsscale{1.25}
	\plotone{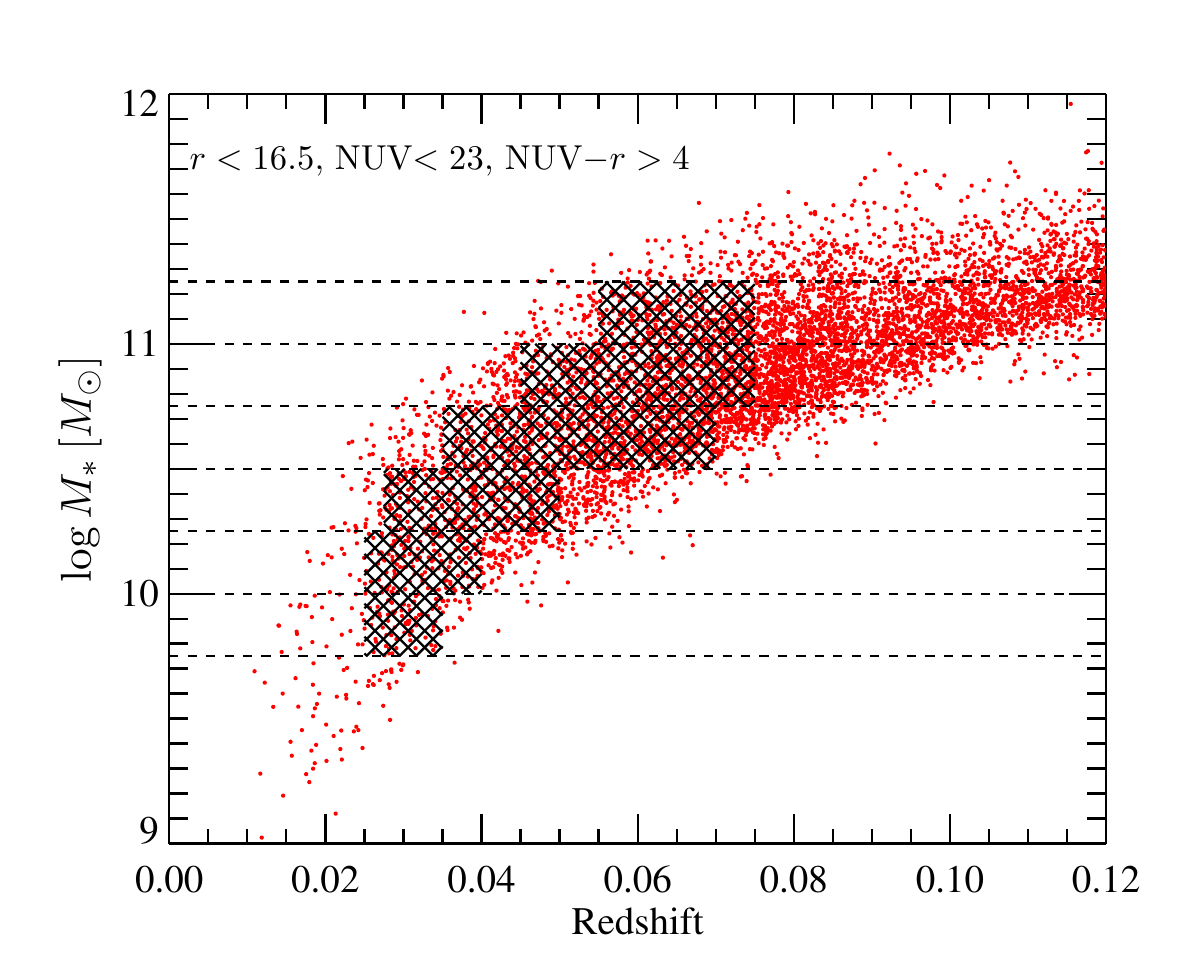}
	
	\caption{Stellar mass vs.~redshift for galaxies in our sample with $r < 16.5$, NUV\,$<23$, and \nuvr$>4$ (red points). The magnitude limits ensure that we are not excluding the reddest objects from the sample (Section \ref{sample}). The horizontal dashed lines indicate the mass bins used in defining our sample. The hatched regions indicate the redshift ranges used to define our ``volume-limited sample.'' This approach maximizes the number of galaxies in each mass bin. Redshift limits and number counts are provided in Table \ref{bins_table}. The size of the SDSS PSF limits our useable redshift range to $z<0.075$ (see Appendix). }

	\label{complete_plot}
\end{figure}

To construct a sample that does not exclude the reddest galaxies yet contains as many objects as possible, we select our final sample in the following way. As can be seen in Figure \ref{cmd_plot}, the upper envelope of the red sequence lies around \nuvr$\approx 6.5$. Given that the NUV magnitude limit is 23, this implies that, to ensure detection of the reddest objects, galaxies must have $r < 16.5$. Figure \ref{complete_plot} shows stellar mass vs.~redshift for galaxies with $r<16.5$ and \nuvr$>4$. Indicated in the figure are the six mass bins used throughout this paper. Each bin is 0.25 dex wide, and we consider the interval $9.75 < \log M_*/M_\odot < 11.25$. For each mass bin, a volume-limited sample is constructed by defining the minimum and maximum redshifts between which red galaxies are detected. Note that the volumes chosen are different for each mass bin. Hereafter, we refer to this sample of 2361 galaxies as the ``volume-limited sample,'' even though it is not formally restricted to a single volume for all galaxies. For reference, Table \ref{bins_table} lists the mass and redshift ranges defining the volume-limited sample, as well as the numbers of galaxies in each bin. The fraction of blue, green, and red galaxies in each mass bin are also provided. Throughout this paper, blue galaxies are defined to have \nuvr $<4$, green galaxies have $4<\ $\nuvr $<5$, and red galaxies have \nuvr $>5$. 

Our use of the SDSS spectroscopic sample can, in principle, bias our sample against specific types of galaxies. Fiber collisions exclude $\sim6\%$ of photometrically detected objects \citep{strauss02}. These excluded galaxies tend to be found in very dense environments or in close pairs. We do not expect the loss of these few objects to alter our conclusions. The sample may also be biased against low surface brightness galaxies. To test this, we identified low surface brightness galaxies using the criteria in \citet{galaz11}. Within the redshift limits of our volume-limited sample, we find that we are complete in low surface brightness galaxies for each mass bin. Such objects comprise only 1.6\% of the volume-limited sample overall and are not expected to affect our conclusions.  




\begin{deluxetable*}{cccccc} 
\tablecaption{Mass and Redshift Ranges and Galaxy Number Counts \label{bins_table}}

\tablehead{\colhead{Mass Range} & \colhead{Redshift Range} &
			\colhead{$N_\mathrm{blue}$\tablenotemark{a}} & 
			\colhead{$N_\mathrm{green}$\tablenotemark{b}} & 
			\colhead{$N_\mathrm{red}$\tablenotemark{c}} &
			\colhead{$N_\mathrm{tot}$} \\
			\colhead{($\log M_*/M_\odot$)} }

\startdata
$[9.75, 10.00]$ & $[0.025, 0.035]$& 53 (70\%) & 10 (13\%) & 13 (17\%)& 76 \\
$[10.00, 10.25]$ & $[0.025, 0.04]$& 91 (55\%) & 21 (13\%) & 52 (32\%)&164 \\
$[10.25, 10.50]$ & $[0.0275, 0.05]$& 134 (42\%)&45 (14\%)&140 (44\%)& 319 \\
$[10.50, 10.75]$ & $[0.035, 0.07]$& 265 (33\%)& 112 (14\%)& 423 (53\%)& 800 \\
$[10.75, 11.00]$ & $[0.045, 0.075]$& 184 (27\%)& 112 (16\%)& 392 (57\%)& 688 \\
$[11.00, 11.25]$ & $[0.055, 0.075]$& 38 (12\%)& 49 (16\%)& 227 (72\%)& 314 \\
 & & \textbf{765 (32\%)} & \textbf{349 (15\%)} & \textbf{1247 (53\%)} & \textbf{2361}
\enddata

\tablenotetext{a}{Galaxies with \nuvr$<4$.}
\tablenotetext{b}{Galaxies with $4<\ $\nuvr$<5$.}
\tablenotetext{c}{Galaxies with \nuvr$>5$.}

\end{deluxetable*}


\subsection{$M/L$ Relation and Mass Surface Density Profiles} \label{sdprof}

\begin{figure}
	\epsscale{1.25}
	\plotone{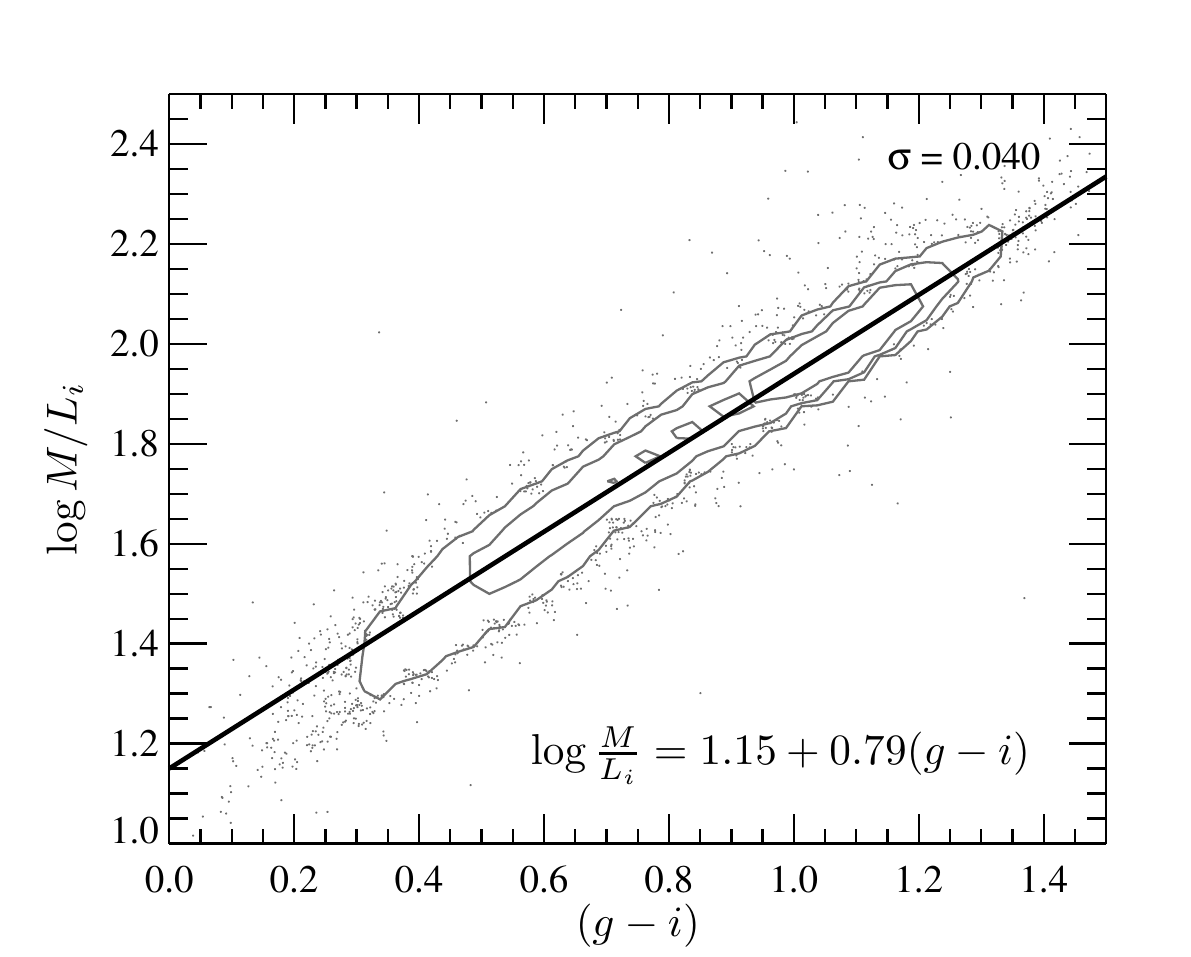}
	
	\caption{Integrated stellar mass-to-light ratio $M/L_i$ ($i$-band luminosity) vs.~rest-frame $g-i$ color for SDSS galaxies in our sample (gray contours and points). Stellar masses are from the MPA/JHU catalog. The best-fit linear relation is plotted, $\log M/L_i=(1.15^{+0.11}_{-0.07})+(0.79^{+0.07}_{-0.10})(g-i)$. Confidence limits for the slope and zeropoint were computed by bootstrap resampling. This relation is used to convert $i$-band surface brightness profiles into mass surface density profiles. The $1\sigma$ scatter about the relation (0.04 dex) is indicated in the upper right.}

	\label{ml_plot}
\end{figure}

A major focus of this paper is studying the buildup of stellar mass in the inner regions of galaxies and its relation to quenching. To trace the mass distribution and its evolution, we compute stellar mass surface density profiles for each galaxy using the surface brightness profiles described in Section \ref{parameters}. To do so requires knowledge of the stellar mass-to-light ratio, $M/L$, to convert surface brightness into mass surface density. We calibrate a relation between $M/L_i$ (i.e., stellar mass divided by $i$-band luminosity) and rest-frame $g-i$ color using the integrated photometry and stellar masses of our sample. Figure \ref{ml_plot} presents the derived calibration, $\log M/L_i=(1.15^{+0.11}_{-0.07})+(0.79^{+0.07}_{-0.10})(g-i)$. The calibration is derived from a linear least-squares fit to the data, incorporating errors in the stellar mass estimates (typically $\sim0.1$ dex; photometric errors are a few percent at most). The uncertainties in the slope and zeropoint are 16th and 84th percentile confidence limits obtained by bootstrap resampling the data points within their errors and recalculating the fit. This relation is consistent, within the uncertainties, with the $M/L_i$ calibration determined by \citet{taylor11} for a larger sample of SDSS galaxies. This $M/L_i$ relation is used in conjunction with the interpolated surface brightness profiles to calculate mass density profiles. The $1\sigma$ scatter about the relation is 0.04 dex, which is taken to be the uncertainty in $M/L_i$ when calculating errors later on. 

A potential concern is how variations in stellar population parameters (e.g., SF history, dust, metallicity, initial mass function) affect our derived $M/L$ relation. In particular, such variations may cause galaxies to scatter off the linear relation, leading to an incorrect estimate of the mass density. However, broadly speaking, such variations conspire together to shift galaxies \emph{along} the relation rather than off it \citep[e.g.,][]{bell01,szomoru13}.

To verify empirically that metallicity does not bias $M/L$, we measured the $M/L$ relation using masses and colors determined within the SDSS 3\arcsec\  fiber. Since the fiber subtends the inner regions of each galaxy ($\la 3$ kpc), we can probe regions with higher metallicities and see if they are located off the relation in Figure \ref{ml_plot}. We find that the fiber-based $M/L$ relation lies on top of the relation in Figure \ref{ml_plot}, with slightly smaller scatter (0.035 dex). No systematic offset is seen, suggesting that metallicity effects do not introduce a systematic bias in our measurements of $M/L$. We have also verified that low surface brightness galaxies follow the main relation and do not lie systematically off it.

Variations in the initial mass function (IMF) can also result in inaccurate estimates of $M/L$. To first order, changing the IMF shifts the zeropoint of the $M/L$ relation but preserves its slope \citep[e.g.,][]{bell01}. If the IMF varies systematically with, e.g., galaxy luminosity, our use of a single calibration would be inappropriate. Indeed, increasing evidence points to systematic variations in the IMF, such that steeper IMFs (more bottom-heavy) are inferred in (early-type) galaxies with higher velocity dispersions \citep[e.g.,][]{cappellari12,conroy12,labarbera13}. If this trend applies to our sample, then our estimates of $M/L$ would be systematically underestimated for more massive and redder galaxies. Fortunately, this effect works in our favor by \emph{increasing} the measured difference in inner stellar surface mass densities between red and blue galaxies, strengthening our claim that quenched galaxies have higher inner mass densities (see Section \ref{sd1_nuvr}). Thus, the changes in our resulting mass density profiles may, in fact, be lower limits to the actual mass distribution changes in quenched galaxies.

\section{THE RELATION BETWEEN INNER SURFACE MASS DENSITY AND COLOR} \label{sd1_nuvr}

Following in the footsteps of \citet{cheung12}, which presented evidence that quenching is linked with \emph{inner stellar mass density} in $z\sim0.8$ galaxies, we explore the link between color and inner mass density with our nearer sample. The average surface mass density within a circular aperture of radius 1 kpc, denoted as \sphy, is computed by directly integrating the mass profiles from the innermost point out to $R=1$ kpc. Our analysis is limited to galaxies with redshift $z<0.075$, where the HWHM of the SDSS PSF (0\farcs7) is comparable to the smallest aperture corresponding to 1 kpc (0\farcs68). The Appendix discusses this issue in more detail.

\begin{figure*}
	\epsscale{1.}
	\plotone{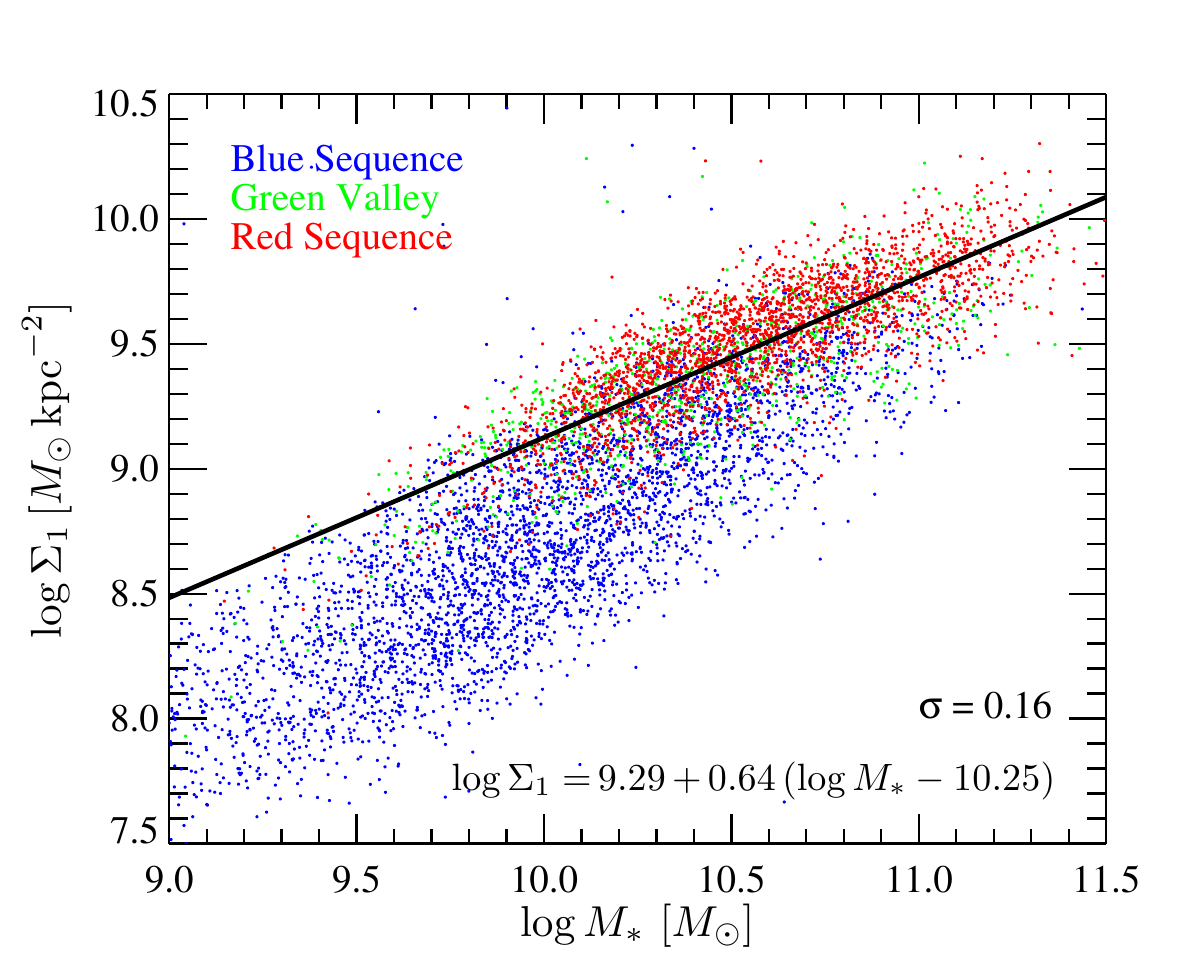}
	
	\caption{\sphy\ vs.~stellar mass for galaxies with $0.005<z<0.075$. Points are color-coded according to \nuvr color. A strong trend between both parameters is seen for green and red galaxies, while blue galaxies lie below the ridge line of the relation. The line is a least-squares fit to the green valley and red sequence galaxies only (\nuvr$>4$), $\log \Sigma_1=(9.29^{+0.10}_{-0.10})+(0.64^{+0.23}_{-0.20})(\log M_*-10.25)$. The quoted confidence limits were obtained using bootstrap resampling of the data. The $1\sigma$ vertical scatter of green and red galaxies about the fit is indicated in the figure (0.16 dex). The strong trend with stellar mass challenges the notion of a fixed, mass-independent threshold surface mass density above which galaxies can quench.} 
	
	\label{sd1_mass_plot}
\end{figure*}

Figure \ref{sd1_mass_plot} shows the distribution of \sphy\ as a function of stellar mass and \nuvr color for galaxies with $0.005<z<0.075$. It is clear that green valley and red sequence galaxies trace out a well-defined, power-law relation. A least-squares fit to only the green and red galaxies is included in the plot, incorporating errors in \sphy. The best-fit relation obtained is
\begin{equation}
\log \Sigma_1=(9.29^{+0.10}_{-0.10})+(0.64^{+0.23}_{-0.20})(\log M_*-10.25),
\label{sd1_mass_eqn}
\end{equation}
with a $1\sigma$ scatter for green and red galaxies about the relation of 0.16 dex. The units are $M_\odot$ for $M_*$ and $M_\odot\, \mathrm{kpc}^{-2}$ for \sphy. The errors are calculated using bootstrap resampling of the data and represent the 16th and 84th percentile confidence limits. This correlation between \sphy\ and stellar mass implies that the threshold value of \sphy\ above which galaxies are seen to quench grows with stellar mass. It thus challenges the notion of a \emph{fixed, mass-independent} surface density threshold \citep[c.f.,][]{kauffmann06,franx08}.   

\begin{figure}
	\centerline{\includegraphics[scale=0.75,bb=30 0 340 428]{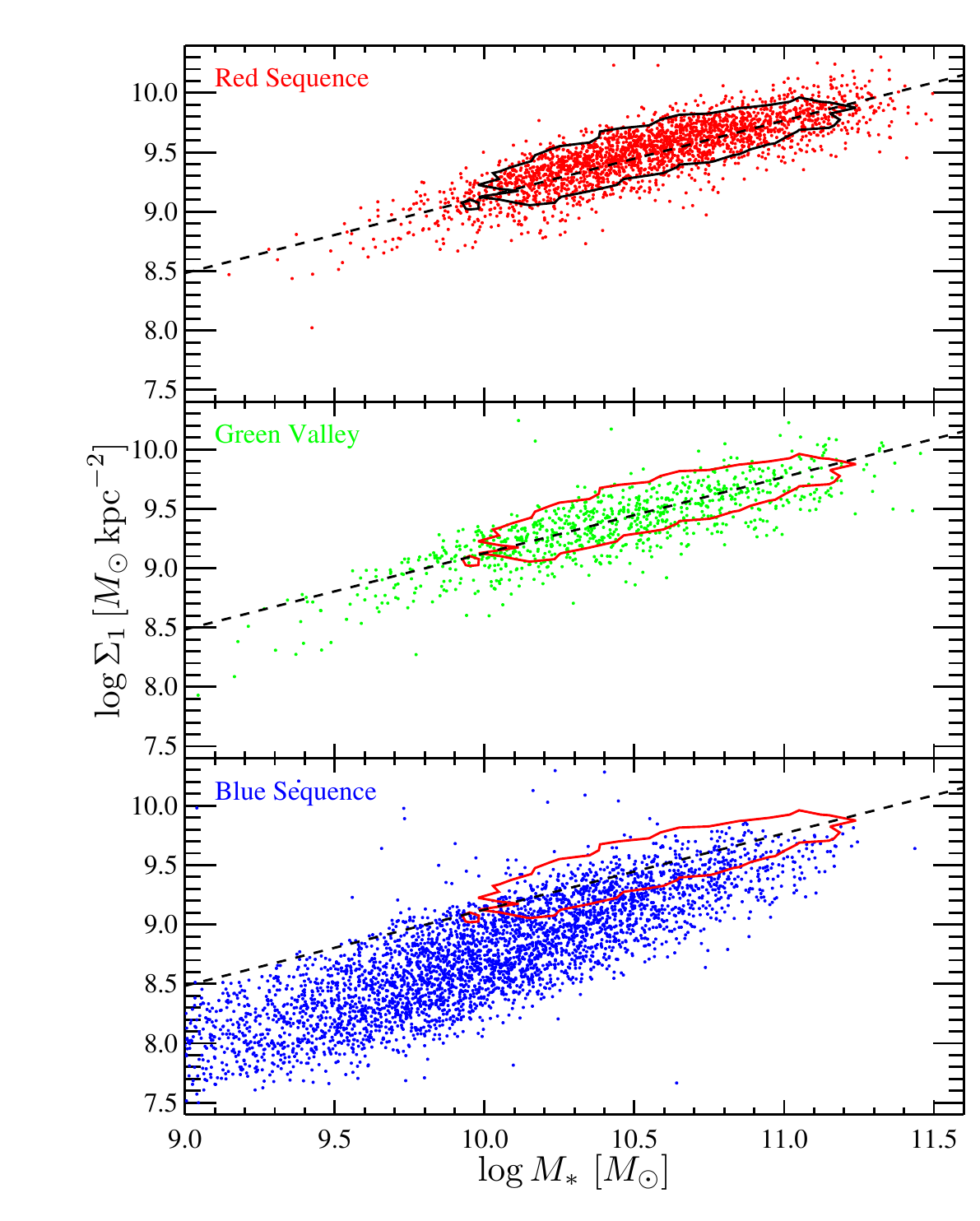}}
	
	\caption{\sphy\ vs.~stellar mass for galaxies with $0.005<z<0.075$. Each panel includes only galaxies in the blue sequence, green valley, and red sequence (\emph{bottom to top}). The black contour defined by the red galaxies in the top panel is repeated in the other panels in red as a visual aid. The best-fit line from Figure \ref{sd1_mass_plot} is also shown in each panel. In general, blue galaxies have lower \sphy\ than quenched galaxies at all stellar masses, showing that quenching is linked with an increase in \sphy. Moreover, the blue galaxies within the contour suggest that a dense bulge is necessary but not sufficient to quench SF. }

	\label{sd1_mass_3slice_plot}
\end{figure}

Figure \ref{sd1_mass_plot} offers a new way to trace the structural changes in a galaxy as it evolves from star-forming to quiescent. Since blue galaxies have systematically lower \sphy\ than green and red galaxies, Figure \ref{sd1_mass_plot} implies that a galaxy increases its \sphy\ as it evolves from blue to red. This is more clearly demonstrated in Figure \ref{sd1_mass_3slice_plot}, which plots the mass-\sphy\ relation for blue, green and red galaxies separately. At all stellar masses, blue galaxies (on average) fall \emph{below} the best-fit relation (Equation \ref{sd1_mass_eqn}) and the locus of points occupied by quenched galaxies (indicated by the red contour). Thus quenching of star formation is accompanied by an increase in \sphy. The implications of this link are explored further in Section \ref{discussion}.  

\begin{figure*}
	\epsscale{1.}
	\plotone{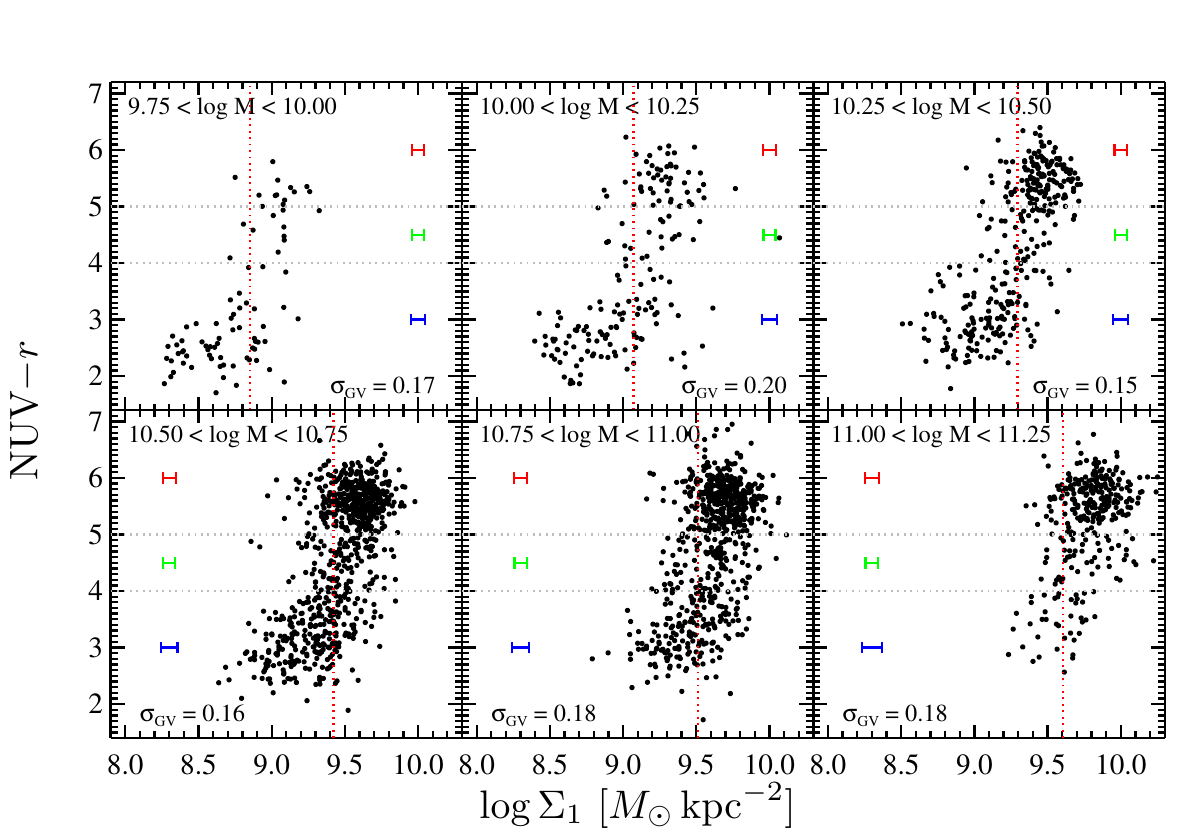}
	
	\caption{\nuvr vs.~\sphy, plotted for the volume-limited sample in six stellar mass bins. Values of \sphy\ are computed from the mass density profiles discussed in Section \ref{sdprof}. The error bars indicate the median error in \sphy\ for blue, green, and red galaxies in each mass bin. Dotted gray lines indicate the division between blue, green, and red galaxies. The ``hook''-shaped distribution in seen in nearly every mass bin. The horizontal scatter in dex of the distribution in the green valley is indicated at the bottom of each panel. Assuming that galaxies evolve at fixed mass, this suggests that galaxies in the blue sequence build up their bulges (i.e., increase \sphy) and then quench and redden. The near-vertical distribution of green and red galaxies suggests that inner bulge buildup does not continue (much) once a galaxy leaves the blue sequence. In addition, the distribution ``marches'' toward higher \sphy\ with increasing stellar mass. Dotted red lines indicate the value of \sphy\ above which 80\% of the red galaxies lie in each mass bin. } 

	\label{nuvr_sd1_plot}
\end{figure*}

A major advantage of our sample is that we have sufficient numbers to investigate the mass dependence of this trend in finer detail. Figure \ref{nuvr_sd1_plot} shows the relation between \sphy\ and \nuvr color in six mass bins 0.25 dex wide ranging from $\log M_*/M_\odot=9.75$ to $\log M_*/M_\odot=11.25$. Dividing the sample into narrow mass bins uncovers systematic trends that are smeared out when examining the whole sample together. In particular, a striking regularity is seen in each mass bin: blue galaxies with \nuvr$< 4$ span a relatively larger range in \sphy, compared to galaxies in the green valley and red sequence (i.e., \nuvr$>4$), which span a \emph{narrower} range in \sphy. The general shape of the distribution resembles a hook, and remains roughly the same across all mass bins. The typical spread in \sphy\ for blue galaxies is $\sim0.7$~dex, while in the green valley and red sequence the spread is smaller, $\sim0.4$~dex.

Since it is known that galaxies evolve from blue to red \citep[e.g.,][]{bell04,faber07,martin07}, it is tempting to interpret the loci seen in Figure \ref{nuvr_sd1_plot} as possible evolutionary tracks, with galaxies evolving roughly horizontally through the blue cloud and then upward onto the red sequence \emph{at fixed stellar mass}. To determine if this is plausible, we divide the proposed path into two segments: (1) evolution through the blue cloud and (2) quenching through the green valley onto the red sequence. For each segment, we compare an appropriately defined ``evolutionary'' timescale with an estimate of the mass doubling time. If both evolutionary timescales are shorter than the mass doubling time, then our claim that galaxies evolve at fixed mass is reasonable.

We first examine the second evolutionary segment by comparing the quenching timescale with the mass doubling timescale. In the local universe, the quenching time (i.e., the time to move across the UV-optical green valley) is of the order 1 Gyr or less \citep{martin07}. The mass growth timescale can be estimated using the results of e.g., \citet{wechsler02,mcbride09,dekel13}. In particular, these authors find that the mass growth of a dark matter halo can be parameterized as an exponential: $M(z)=M_0e^{-\alpha z}$, with the best-fitting $\alpha\approx0.7$. Using this formula, we find that the time to double the halo mass is $\approx8$ Gyr (i.e., the time elapsed since $z\approx1$). A similar doubling time for the stellar mass is found in semi-analytic models of galaxy growth \citep[][]{behroozi13}. Since this time is much longer than the quenching timescale, we can safely assume that galaxies do not significantly increase their mass as they quench, i.e., they remain in the same mass bin as they leave the blue cloud, move through the green valley, and arrive on the red sequence. This picture neglects major mergers, but they are expected to be rare at low redshifts \citep{lotz11} and an insignificant contribution to the overall mass growth \citep{moustakas13}. In summary, since the quenching timescale is much shorter than the mass doubling time, galaxies will, on average, evolve through the green valley onto the red sequence \emph{at fixed stellar mass}. 

Some of the green valley objects in Figure \ref{nuvr_sd1_plot} are prime candidates for rapid SF quenching, namely, post-starburst galaxies. While we do not explicitly identify them in this work (since we do not discuss spectral information), other studies find that post-starbursts lie in the green valley and have compact structure (high \sphy), consistent with the evolutionary tracks in Figure \ref{nuvr_sd1_plot} \citep[][H.~Yesuf, in preparation]{wong12,mendel13}.

We now turn to the first segment of the proposed path and ask if blue galaxies can increase \sphy\ without a significant (factor of 2) increase in stellar mass. We can appeal to another timescale argument to verify this. We assume a mass doubling time of $\sim8$ Gyr, as above. If the growth in \sphy\ is due to e.g., violent disk instabilities, the timescale for clumps in the disk to fall toward the center is of order tens of dynamical times \citep[i.e., a few Gyr;][]{dekel09}. However, violent disk instabilities are believed to be more common at high redshift and are unlikely to be the main driver of bulge buildup in our galaxies. On the other hand, secular evolution (via, e.g., bar-driven processes) is believed to be a key player at late times and can also increase the inner mass density on timescales of a few Gyr \citep{kormendy04}. While these bulge buildup timescales are smaller than the mass doubling time, we caution that these estimates are somewhat uncertain. Despite this caveat, it is nevertheless plausible that galaxies can build up their inner regions while maintaining roughly constant stellar mass since $z\sim1$. Thus our interpretation of each panel of Figure \ref{nuvr_sd1_plot} as an evolutionary track is reasonable.     

Our interpretation is also consistent with the data themselves. Specifically, the distribution of points in Figure \ref{nuvr_sd1_plot} (especially at lower masses) forms a ``hook'' shape, with some blue galaxies having lower \sphy\ than green and red galaxies of the same mass. The lack of quenched galaxies with \emph{low} \sphy\ in each mass bin is strong evidence that blue galaxies cannot simply fade onto the red sequence without increasing their inner mass density. If they could, we would expect to see the scatter in \sphy\ for green and red galaxies to be comparable to the scatter in \sphy\ for blue galaxies. This is not seen in Figure \ref{nuvr_sd1_plot}. Our interpretation is consistent with \citet{cheung12}, who find that the inner mass concentration of galaxies at $z\sim0.8$ must increase in order for galaxies to quench. Our result moreover shows that this mass rearrangement occurs \emph{at fixed stellar mass}. We note that the few low surface brightness galaxies in our sample lie in the low-\sphy\ tail of the blue cloud in Figure \ref{nuvr_sd1_plot}, along with the many other normal galaxies with low \sphy. Thus such objects do not bias our interpretation of the figure.


In addition to the hook-shaped distribution, Figure \ref{nuvr_sd1_plot} reveals other trends. One is the ``disappearance'' of blue sequence galaxies in higher mass bins (Table \ref{bins_table}). This is not surprising because the blue sequence is tilted in color-mass space (see Figure \ref{cmd_plot}), with the bluest objects found only at low stellar masses. Put another way, the ratio of blue galaxies to red galaxies decreases with stellar mass (Table \ref{bins_table}).  If galaxies do evolve as we have described, then Figure \ref{nuvr_sd1_plot} also implies that such evolution is more complete at higher masses. That is, the dearth of massive blue galaxies is consistent with more massive galaxies completing their evolution onto the red sequence at earlier times. This is consistent with downsizing and also the fact that more massive spheroidal galaxies are found to possess more evolved stellar populations \citep[e.g.,][]{thomas05,cattaneo08,graves10a,barro13,moustakas13}. Our new results imply that more massive galaxies are more dynamically evolved as well. 

\section{SURFACE BRIGHTNESS AND MASS DENSITY PROFILES} \label{profiles}

In this section we examine the mass profiles for further insight into the global mass distribution among galaxies of different colors. Our main goal is to compare the profiles of galaxies in the blue sequence, green valley, and red sequence at fixed stellar mass to determine how they differ and if they shed light on a plausible evolutionary path. We also highlight important differences between mass-weighted radii and light-weighted radii as measures of galaxy size.  

\subsection{Median Surface Brightness and Mass Density Profiles as a Function of Color}

\begin{figure*}
	\centerline{\includegraphics[scale = 1.1, bb = 25 0 340 275]{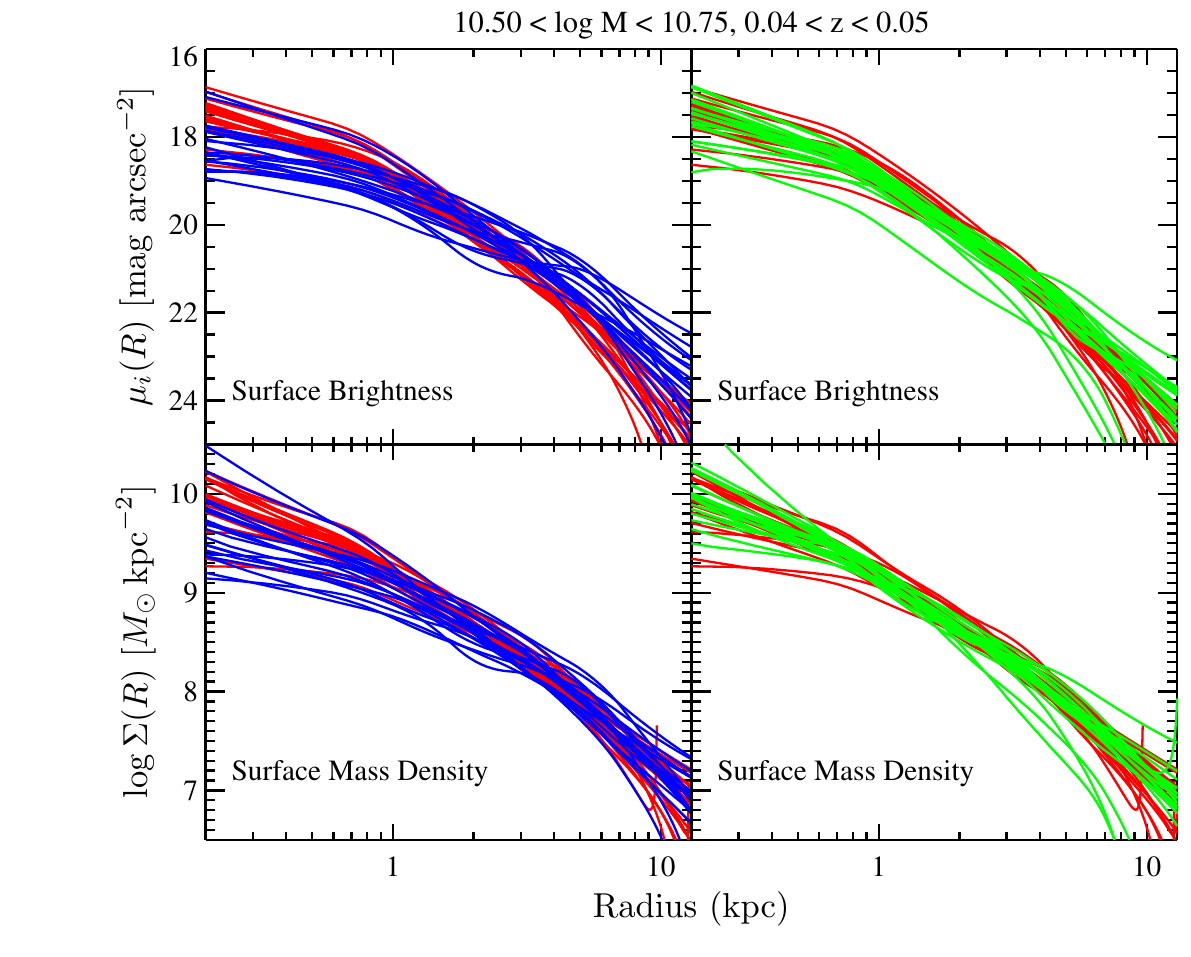}}
	
	\caption{\emph{Top row:} SDSS $i$-band surface brightness profiles for 20 representative blue, green, and red galaxies with stellar mass $10.5<\log M_*/M_\odot < 10.75$ and redshift $0.04<z<0.05$. \emph{Bottom row:} Stellar surface mass density profiles for the same galaxies (derived from the $i$-band profiles). While blue galaxies have brighter surface brightness profiles at large radii, their outer mass profiles are very similar to those of green and red galaxies. In addition, blue galaxies have lower inner mass densities compared to the (quenched) green and red objects.}

	\label{prof_slice_plot}
\end{figure*}

Before presenting profiles for the whole sample, we show, in the top row of Figure \ref{prof_slice_plot}, $i$-band surface brightness (SB) profiles, $\mu_i(R)$, for a random sample of 20 blue, green, and red galaxies in a single mass and redshift slice. To facilitate comparison, we overplot blue and red galaxies in the left column and green and red galaxies in the right column. Starting with the upper-left panel, we see that blue galaxies have systematically brighter outer regions than red galaxies but have fainter inner regions. In the upper-right panel, we see that green and red galaxies have rather similar $i$-band profile shapes. Taken together, these SB profiles show that as galaxies evolve from blue to red (at fixed mass), their outer disks fade and their inner regions get brighter.  

The bottom row of Figure \ref{prof_slice_plot} converts these brightness profiles to surface mass density (SD) profiles, $\Sigma(R)$, using the $M/L$ relation in Figure \ref{ml_plot}. Despite their brighter outer disks, blue galaxies have \emph{nearly identical} outer mass profiles as red galaxies. This can be explained as a consequence of the lower mass-to-light ratios of blue galaxies (i.e., while blue galaxies have bright disks due to young stars, those stars do not dominate the total stellar mass). Despite the similar outer mass profiles, it can be seen that red galaxies have, on average, higher mass densities in the inner regions ($R\la$ a few kpc) than blue objects. The lower-right panel shows that the SD profiles of green and red galaxies are very similar at all radii. 

\begin{figure*}
	\epsscale{1}
	\plotone{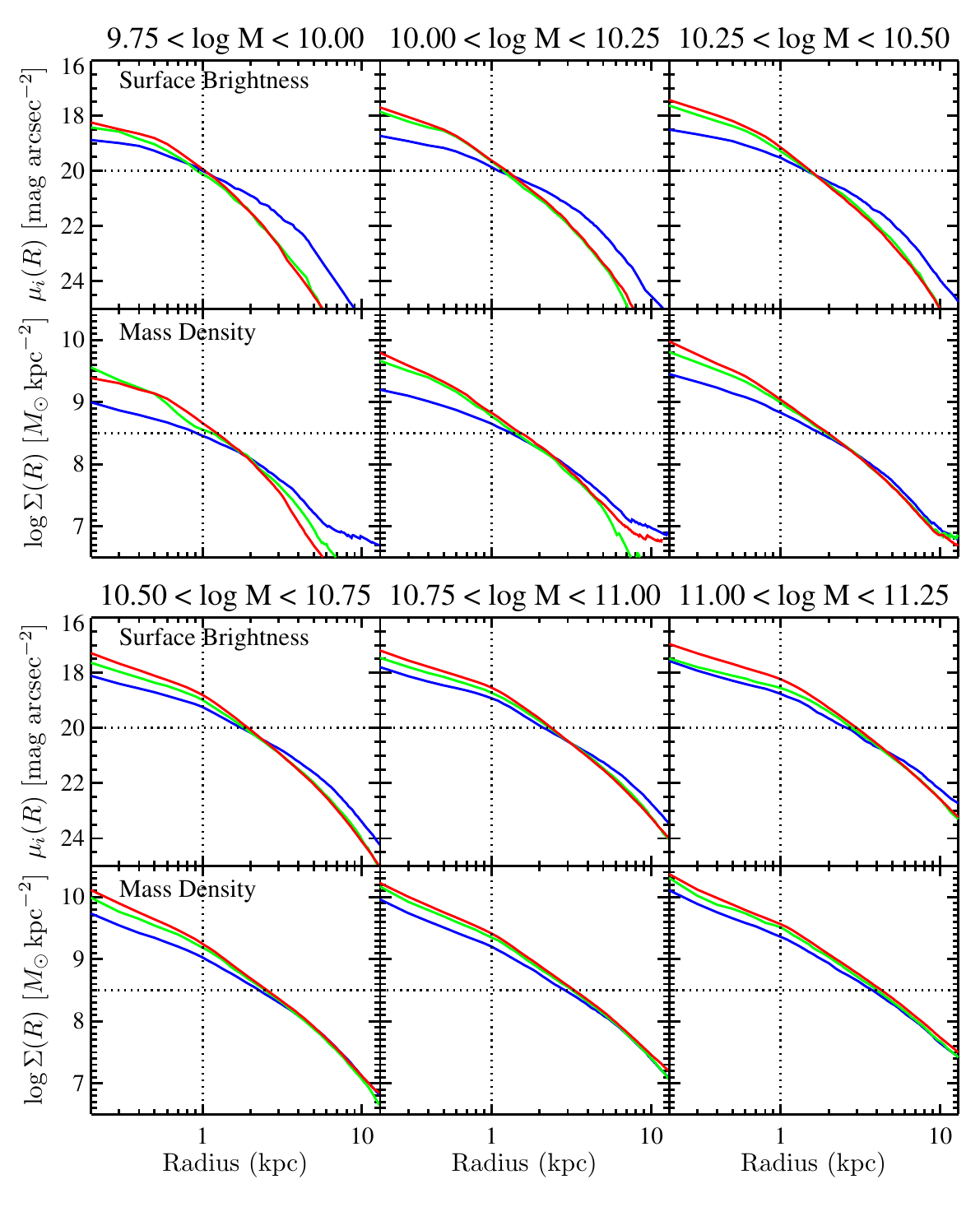}
	
	\caption{Median SDSS $i$-band surface brightness profiles and mass surface density profiles (derived from the $i$-band profiles) for galaxies in six stellar mass bins for the volume-limited sample. Profiles are color-coded to correspond to galaxies in the blue sequence, green valley, and red sequence. While the surface brightness profiles for green and red galaxies are nearly identical, blue galaxies have dimmer centers and brighter outer disks; this effect is more pronounced at lower masses. However, the outer mass density profiles are nearly identical for galaxies of all colors, particularly at higher masses. Dotted lines in each panel are the same and serve to guide the eye.}
	
	\label{sb_sd_prof_plot}
\end{figure*}

We now move beyond a single mass and redshift bin to examine the SB and SD profiles of all galaxies in the volume-limited sample. Figure \ref{sb_sd_prof_plot} presents the median $i$-band SB and SD profiles for galaxies in the sample, divided into the same mass bins used in Figure \ref{nuvr_sd1_plot}. In each mass bin, galaxies are divided into blue, green, and red galaxies based on \nuvr color. The conclusions drawn from Figure \ref{prof_slice_plot} are essentially the same for the whole sample. Beginning with the SB profiles, what is apparent from the figure is that the SB profiles of green and red galaxies are remarkably similar at all radii in all mass bins. This suggests that green and red galaxies are structurally very similar. The second notable aspect is the difference between the SB profiles of the blue and green/red galaxies. In particular, the blue galaxies have significantly brighter outer regions. This is not surprising given the fact that blue galaxies predominately have extended (bright) disks. 

Turning to the SD profiles, inspection of Figure \ref{sb_sd_prof_plot} reveals that the outer parts of \emph{all} galaxies in a mass bin have  similar values of $\Sigma(R)$, especially at higher mass. This stands in contrast to the SB profiles, where we noted that blue galaxies have systematically brighter outer profiles than green and red galaxies. In addition, the mass profiles highlight that the main difference in $\Sigma(R)$ between galaxies of different colors is found in their inner regions ($R\la1$ kpc). At fixed mass, green and red galaxies have inner surface densities systematically larger than blue galaxies by about a factor of 2--3. As mentioned in Section \ref{sdprof}, recently reported IMF variations would \emph{increase} the difference in inner $\Sigma(R)$ between blue and red galaxies.

Taking the view that galaxies evolve from blue to red at fixed mass, we are led to conclude that mass is building up in the inner (bulge-dominated) regions as galaxies evolve. Moreover, most of this buildup occurs \emph{before the galaxy leaves the blue cloud}. This last point is indicated by the observation that the SD profiles of green and red galaxies are nearly identical in their inner regions. If, instead, mass buildup occurred gradually as a galaxy transitioned from blue to red, we would expect to see the green valley mass profiles take on values more intermediate between the blue and red galaxy profiles. This expectation is strengthened by referring back to Figure \ref{nuvr_sd1_plot}. In each mass slice, especially at lower stellar masses where the blue cloud is well-populated, the distribution of galaxies in color-\sphy\ space is consistent with an interpretation in which galaxies increase their inner mass densities while still in the blue cloud. Once they reach a critical value of \sphy, galaxies are able to quench and move into the green valley. Moreover, \sphy\ remains nearly constant after a galaxy is quenched; this manifests itself as the nearly vertical tracks in Figure \ref{nuvr_sd1_plot}.

\begin{figure*}
	\epsscale{1}
	\plotone{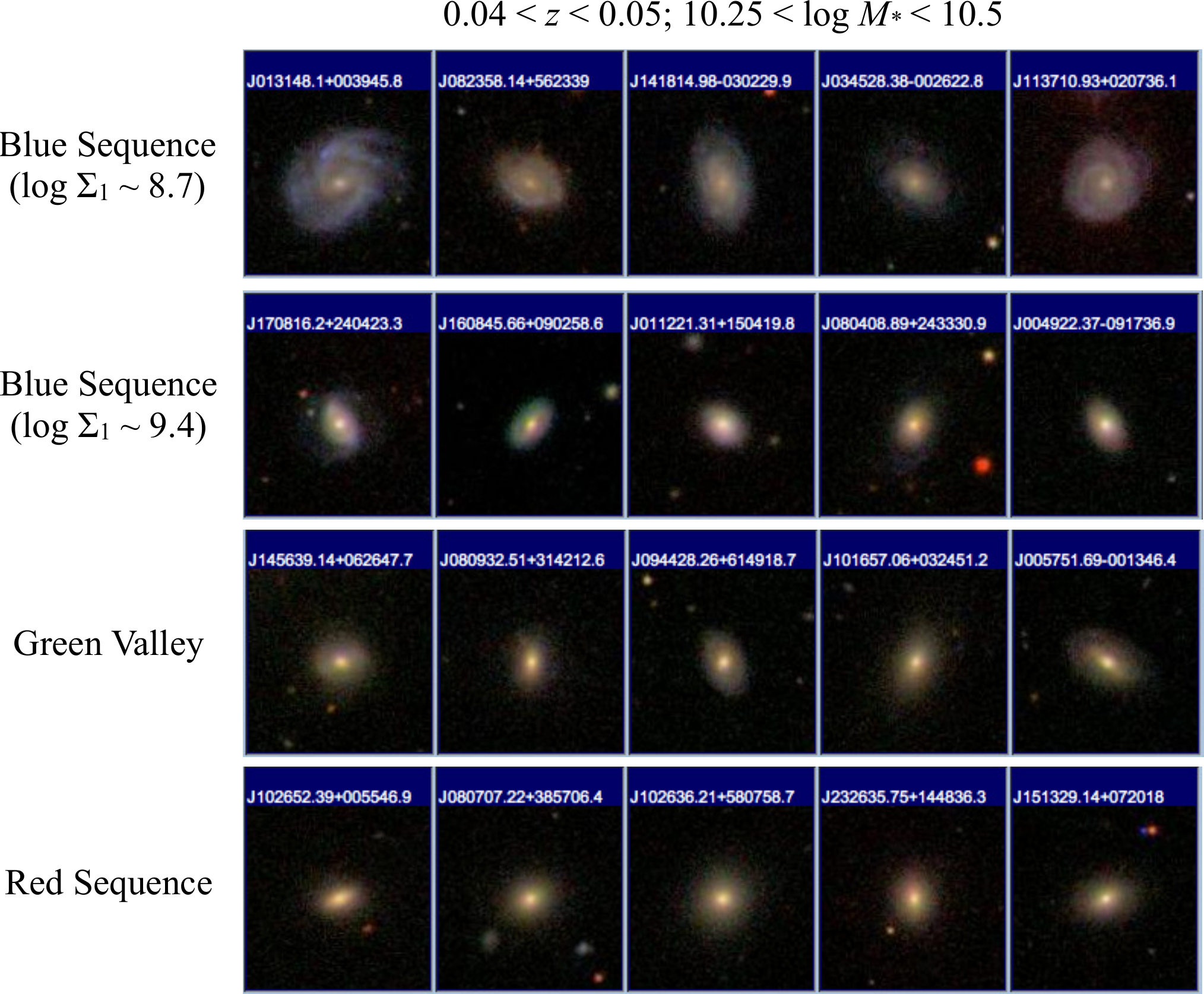}
	
	\caption{Montage of SDSS postage stamps ($\sim25$ kpc on a side) for a selection of galaxies with redshifts $0.04 < z < 0.05$ and stellar masses $10.25 < \log M_*/M_\odot < 10.5$. \emph{Top row:} Blue sequence galaxies with relatively low \sphy\ ($\log$\,\sphy\,$\approx8.7$). \emph{Second row:} Blue sequence galaxies with relatively high \sphy\ ($\log$\,\sphy\,$\approx9.4$). Note the striking differences between even blue sequence galaxies at the same mass. \emph{Third row:} Green valley galaxies. \emph{Bottom row:} Red sequence galaxies. Note that green and red galaxies have similar sizes and more prominent bulges than the blue galaxies in the top row. However, the blue galaxies in the second row bear a strong resemblance to the green and red galaxies. It is plausible that the green and red galaxies are the quenched and faded counterparts of the high-\sphy\ blue galaxies in the second row.}

	\label{montage_plot}
\end{figure*}

Figure \ref{montage_plot} gives a visual sense of the diversity of galaxy morphologies, even at fixed mass. This montage of SDSS color postage stamps shows a random sampling of galaxies with redshifts between 0.04 and 0.05 and stellar masses $10.25 < \log M_*/M_\odot < 10.5$. The narrow redshift slice was chosen to ensure that the images have comparable physical extent (25 kpc). The top row shows blue galaxies with low \sphy. As can be seen, these galaxies have extended blue disks and weak central bulges. Given their dim centers, low surface brightness galaxies fall into this category (e.g., the fourth object in the top row). The second row in Figure \ref{montage_plot} displays blue galaxies with higher \sphy. For consistency, both rows show galaxies with \nuvr $<3$. What is striking is the radically different optical sizes and visual morphologies of the galaxies in the second row compared to those in the top row. These high-\sphy\ objects have much smaller disks and more pronounced bulges.  

The third and fourth rows of Figure \ref{montage_plot} show examples of green valley and red sequence galaxies, respectively. Their sizes and morphologies are very similar to the high-\sphy\ blue cloud objects in the second row. This lends further support to a scenario where nearly all of the bulge buildup takes place while galaxies are still in the blue cloud. Thus, the mass rearrangement must be essentially complete \emph{before} galaxies move into the green valley. 

\subsection{Light-Weighted vs.~Mass-Weighted Radii}

Figure \ref{sb_sd_prof_plot} shows that the SB profiles show systematic differences from the SD profiles. For example, at fixed mass, blue galaxies have brighter SB profiles at large radii than quenched objects, yet their outer mass profiles are nearly identical to those of quenched galaxies. This implies that apparent differences in e.g., sizes of blue and red galaxies are likely exaggerated simply because sizes are measured using SB profiles, which are affected by bright blue stars in the disks of star-forming galaxies. This effect has been called ``outshining'' in some contexts \citep{reddy12,wuyts12}, whereby light from recently formed stars biases measurements of radii, stellar population ages, or other parameters.

\begin{figure*}
	\epsscale{0.9}
	\plotone{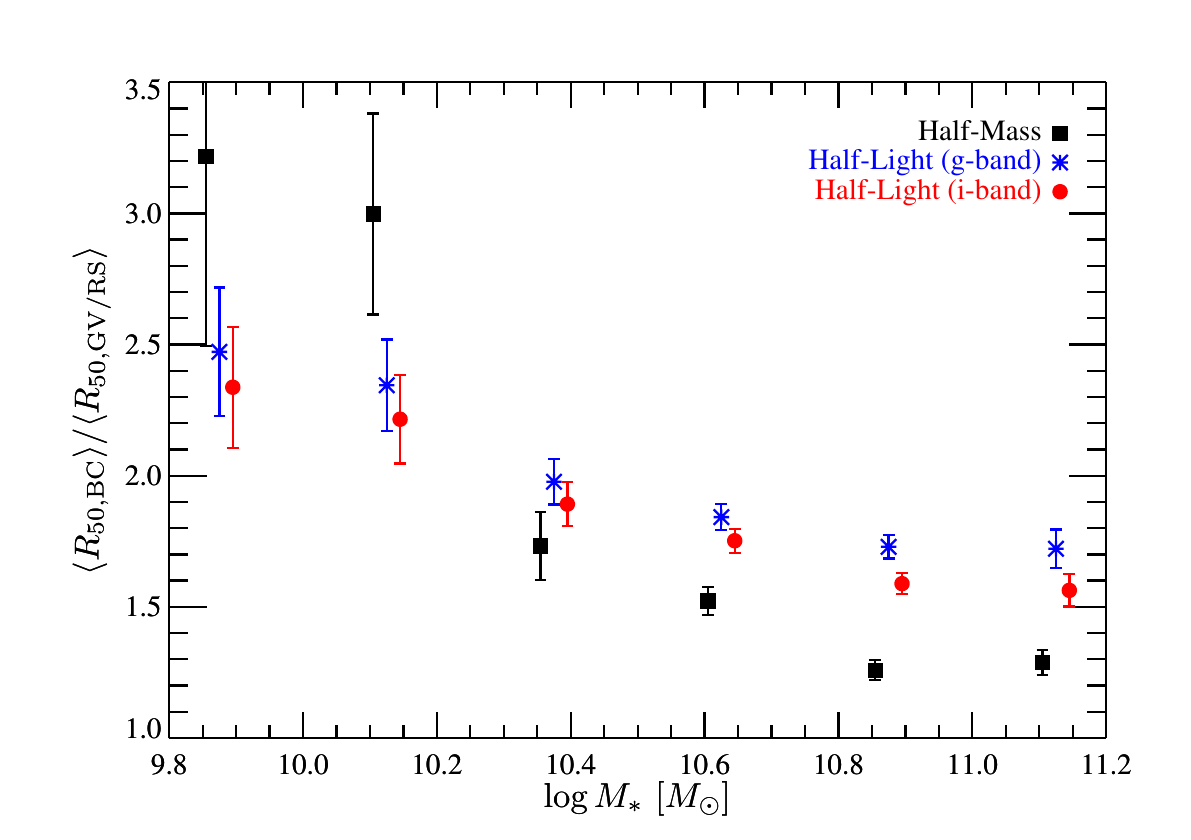}
	
	\caption{Ratio of mean galaxy radius of blue cloud galaxies to quenched galaxies, $\langle R_{50,\mathrm{BC}}\rangle/\langle R_{50,\mathrm{GV/RS}}\rangle$, as a function of stellar mass. Each symbol represents a different definition of radius:~$g$-band half-light radius (blue stars), $i$-band half-light radius (red circles), and half-mass radius (black squares). While at lower masses all three radii give consistent ratios, light-weighted radii exaggerate the difference in size between star-forming and quenched galaxies at higher masses. Individual points have been offset horizontally for clarity. Error bars indicate the uncertainty in the mean ratio, and are large at lower masses due to the small number of objects in those bins. }

	\label{radii_ratio_plot}
\end{figure*}

Figure \ref{radii_ratio_plot} highlights the effects of outshining by comparing measures of galaxy size using light-weighted radii and mass-weighted radii. The ratio of the mean radius for blue galaxies to the mean radius for green and red galaxies, $\langle R_{50,\mathrm{BC}}\rangle /\langle R_{50,\mathrm{GV/RS}}\rangle$, is plotted as a function of stellar mass. Three radius definitions are used: $g$-band half-light radius, $i$-band half-light radius, and half-mass radius. Half-light radii are the 50\% Petrosian radii from the SDSS database, and half-mass radii are computed from the SD profiles.

For each radius definition, the general trend is that the size difference between star-forming and quenched galaxies decreases slightly with stellar mass. This trend is consistent with the observed increase in bulge-to-total ratios with stellar mass \citep[e.g.,][]{kauffmann03a}. That is, more massive galaxies, regardless of color, tend to be more bulge-dominated than lower-mass objects, which reduces the difference in size between blue and red galaxies. What is striking, however, is the comparison between different radius definitions at a fixed mass. In particular, for $M_*\ga 10^{10.25}\, M_\odot$, light-weighted radii significantly exaggerate the difference in size between blue and green/red galaxies, when compared to mass-weighted radii, which ostensibly represent ``true'' sizes. As expected, the effect is slightly more pronounced when using the bluer $g$-band radii, which are more sensitive to younger stars than $i$. That the half-mass size ratio is always larger than unity indicates that quenched galaxies are more compact than their star-forming progenitors when using mass-weighted sizes. This is due, presumably, to the growth of the bulge as galaxies evolve.

Because of outshining, light-based structural parameters are not generally indicative of the underlying stellar mass distribution of a galaxy. \citet{szomoru13} show that mass-weighted radii are $\sim25\%$ smaller than light-weighted radii for massive galaxies out to $z\sim2$. A similar result is seen in \citet{wuyts12} using resolved mass maps of massive, star-forming galaxies out to $z\sim2.5$. Our analysis sharpens these results by separating objects according to both color and mass and discussing how outshining affects star-forming and quiescent galaxies \emph{differentially}. Indeed, we caution that this differential outshining may exaggerate apparent trends in e.g., Sersic index as a function of color. It is well-known that blue galaxies have mainly low Sersic indices ($n\approx1$) while red galaxies predominately have higher values ($n\approx4$). This is in part due to the higher bulge-to-total ratios in red galaxies. However, as the SB profiles show, while it is clear that the bulge component increases in strength from blue to red, the outer disks also \emph{fade} as galaxies redden. These two effects act together to increase the apparent difference in $n$. Our analysis shows that the underlying mass density profile is nearly the same among blue and red galaxies, especially in the outer parts, i.e., the mass density does not ``fade'' as a galaxy reddens. It thus appears that the ``bimodality'' of Sersic values is exaggerated by fitting light profiles rather than mass profiles. A comparison between light-weighted and mass-weighted Sersic indices is urgently needed, but it is beyond the scope of this paper.

\section{COMPARING INNER SURFACE MASS DENSITY WITH VELOCITY DISPERSION} \label{comparison}

Our results thus far have revealed that \sphy\ is correlated with the SF state of galaxies and may serve as a useful diagnostic of quenching. In this section, we examine the use of velocity dispersion as an alternative tracer of quenching, as proposed in \citet{wake12}, and compare it with \sphy. We also discuss how measurements of \sphy\ may offer a new method of examining black hole growth in galaxies.  

\subsection{The Relation Between Velocity Dispersion and Stellar Mass}

\begin{figure*}
	\epsscale{1}
	\plotone{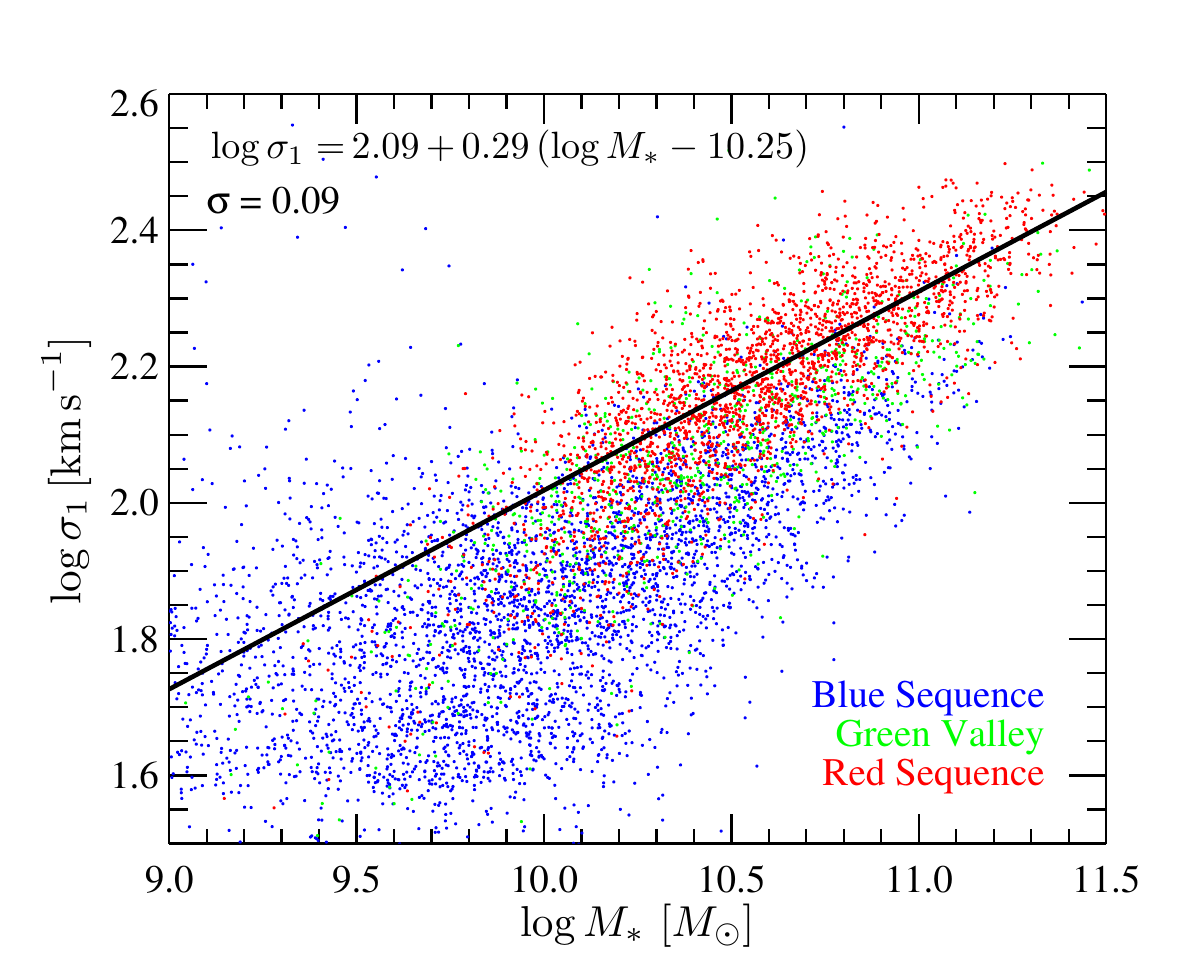}
	
	\caption{Velocity dispersion scaled to a 1-kpc aperture, \vsc, vs.~stellar mass for galaxies with $0.005<z<0.075$. The points are color-coded by \nuvr color. The line is a least-squares fit to green and red galaxies only, $\log\sigma_1 = 2.09^{+0.05}_{-0.08}+(0.29^{+0.13}_{-0.10})(\log M_*-10.25)$. The $1\sigma$ vertical scatter of green and red galaxies about the relation is indicated in the figure (0.09 dex). The strong trend with stellar mass supports the notion of a mass-dependent quenching threshold, previously seen in Figure \ref{sd1_mass_plot}.}

	\label{vdisp_mass_plot}
\end{figure*}

Figure \ref{vdisp_mass_plot} plots velocity dispersion scaled to a circular aperture of radius 1 kpc, \vsc, vs.~stellar mass for galaxies with $0.005<z<0.075$. Points are color-coded according to their \nuvr color. A least-squares fit to green and red galaxies is also plotted: 

\begin{equation}
	\log\sigma_1 = 2.09^{+0.05}_{-0.08}+(0.29^{+0.13}_{-0.10})(\log M_*-10.25),
\end{equation}
with the uncertainties corresponding to the 16th and 84th percentile confidence limits obtained by bootstrap resampling the data within their errors. SDSS fiber velocity dispersions were scaled using the relation in \citet{cappellari06} calibrated using the early-type galaxies in the SAURON survey, $\sigma\propto R^{-0.66}$. The use of \vsc\ facilitates comparison with \sphy.

The distribution of points in Figure \ref{vdisp_mass_plot} is very similar to that seen in Figure \ref{sd1_mass_plot}. In particular, green and red galaxies lie on a well-defined, mass-dependent relation, while blue galaxies scatter below the ridge line. Since \vsc\ is determined spectroscopically, it is a more direct (dynamical) tracer of the total mass distribution (baryonic + dark matter) in the inner regions. This allows us to use \vsc\ as an independent check of our previous result that the inner mass distribution builds up as galaxies evolve from blue to red. The consistent mass dependence of both \sphy\ and \vsc\ imply that this buildup is a robust result and not simply due to possible systematics in converting light to mass when measuring \sphy. 

\subsection{The Relation Between \sphy\ and Velocity Dispersion} \label{sd1_vd}

\begin{figure*}
	\epsscale{1}
	\plotone{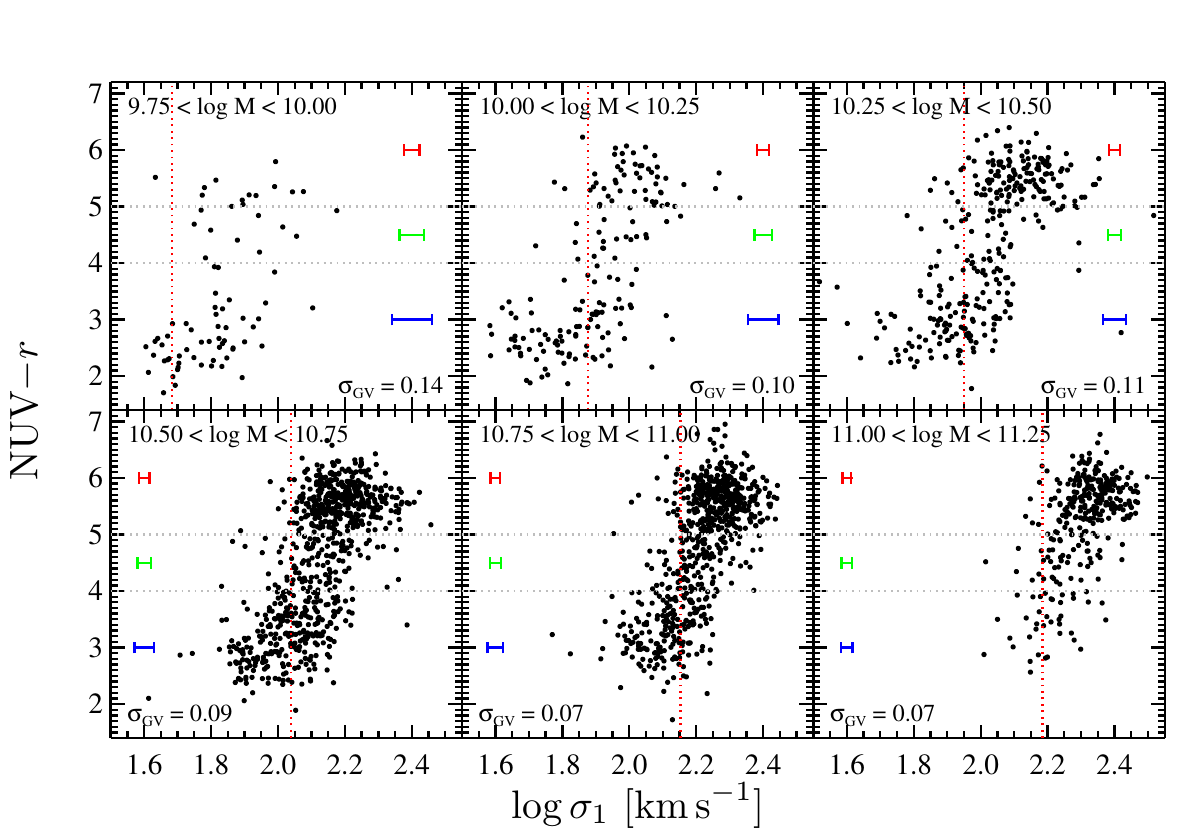}
	
	\caption{\nuvr vs.~\vsc\ in six stellar mass bins for the volume-limited sample. Error bars indicate median errors in \vsc\ for blue, green, and red galaxies in each mass bin. Dotted gray lines indicate the division between blue, green, and red galaxies. The horizontal scatter in dex of the distribution in the green valley is indicated at the bottom of each panel. The dotted red lines indicate the value of \vsc\ above which 80\% of the red galaxies are found in each mass bin. Though a correlation between \vsc\ and \nuvr color is apparent, the scatter and measurement errors in \vsc\ become increasingly severe toward lower masses.}

	\label{nuvr_vdisp_slice_plot}
\end{figure*}

To test recent claims that velocity dispersion is better correlated with color than other structural parameters \citep{wake12}, Figure \ref{nuvr_vdisp_slice_plot} plots \nuvr color against velocity dispersion in six stellar mass bins. This plot can be compared to Figure \ref{nuvr_sd1_plot}. Inspection of Figure \ref{nuvr_vdisp_slice_plot} shows that \vsc\ is fairly well-correlated with \nuvr color in galaxies more massive than $\sim10^{10.25}\,M_\odot$. However, at lower masses, the correlation is less obvious because the distribution shows increased scatter, possibly due to increased measurement errors in \vsc. This is in contrast to the behavior of \sphy, which shows comparatively little scatter in all mass bins (Figure \ref{nuvr_sd1_plot}). Except for this, the two figures are remarkably similar. 

\begin{figure*}
	\epsscale{1}
	\plotone{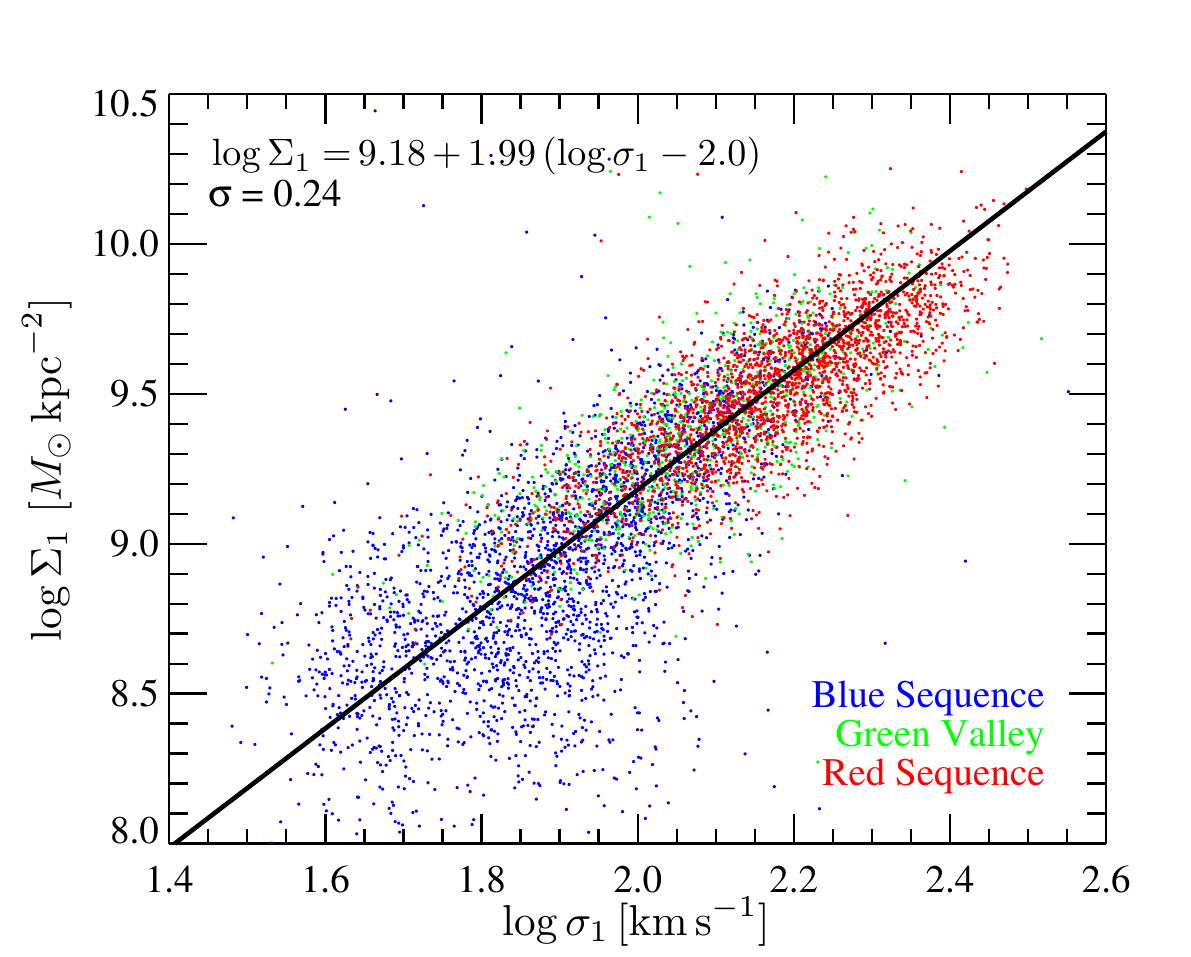}
	
	\caption{\sphy\ vs.~\vsc. Galaxies with $9.75<\log M_*/M_\odot<11.25$ and $0.005<z<0.075$ are plotted. Points are color-coded according to \nuvr color. The line is a least-squares fit to \emph{all} galaxies in the plot incorporating errors in both quantities. There is a strong correlation between the inner surface density and velocity dispersion, $\log \Sigma_1=(9.18^{+0.04}_{-0.05})+(1.99^{+0.22}_{-0.16})(\log \sigma_1-2.0)$. The slope of the relation is consistent with simple dynamical arguments (see text). The existence of such a relation suggests that both quantities are sensitive to bulge buildup. The $1\sigma$ vertical scatter about the relation is indicated at upper left (0.24 dex). The scatter is reduced to 0.18 dex when including only the green and red galaxies.}

	\label{sd1_vdisp_plot}
\end{figure*}

\sphy\ and \vsc\ are compared directly in Figure \ref{sd1_vdisp_plot}. As can be seen, the two quantities are strongly correlated. The line is a least-squares fit of all galaxies, incorporating errors in both \sphy\ and \vsc, 
\begin{equation}
\log \Sigma_1=(9.18^{+0.04}_{-0.05})+(1.99^{+0.22}_{-0.16})(\log \sigma_1-2.0)
\end{equation}
with \sphy\ in $M_\odot\,\mathrm{kpc}^{-2}$ and $\sigma_1$ in $\mathrm{km\, s^{-1}}$. Bootstrap resampling was used to compute uncertainties in the slope and zeropoint. The $1\sigma$ vertical scatter about the relation is 0.24 dex. The scatter is reduced to 0.18 dex when considering only green and red galaxies. Intriguingly, the power-law slope of $\sim2$ between \sphy\ and \vsc\ is consistent with predictions from simple scaling arguments. Neglecting projected mass outside a radius $R$ and assuming homologous structure, the surface density $\Sigma\propto M/R^2$, and the mass is related to velocity dispersion and radius by the virial theorem ($M\propto\sigma^2 R$). Combining these relations we obtain
\begin{equation}
\Sigma\propto\frac{\sigma^2}{R},
\end{equation}
or, assuming all quantities are measured within a constant aperture of 1 kpc,
\begin{equation}
\Sigma_\mathrm{1}\propto\sigma_\mathrm{1}^2.
\end{equation}
Thus simple scaling arguments predict a power-law relation with a slope of 2, which matches the observationally determined slope. We stress that our use of the virial theorem assumes that galaxy bulges are homologous, which is not necessarily the case, considering the large diversity of galaxy morphologies in our sample, and it also neglects contributions to the surface mass density from regions that are outside a 1-kpc radius sphere centered on the nucleus. In spite of this, the consistency between the observed and predicted relations is an intriguing result. 

The tight correlation between \sphy\ and \vsc\ implies that the value of one quantity can be predicted given knowledge of the other (especially for quenched galaxies, where the scatter is smaller). In particular, this relation might be used to estimate velocity dispersions in local galaxies without the need to obtain time-consuming spectra. Given the relative abundance of high-resolution photometry of high-redshift galaxies, the result might even be a useful tool in studying the dynamical properties of more distant objects. Of course, it would first have to be established that this correlation between mass density and velocity dispersion remains valid at higher redshifts.

\begin{figure*}
	\epsscale{1}
	\plotone{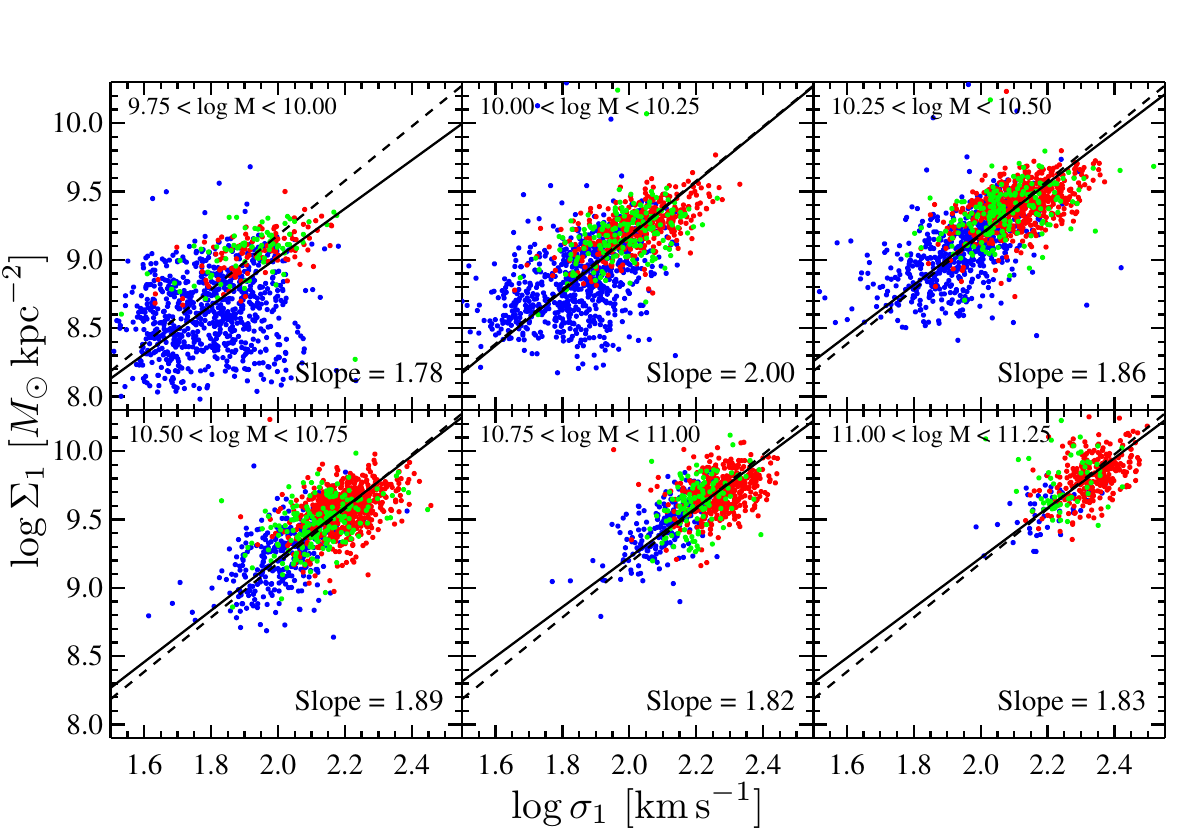}
	
	\caption{\sphy\ vs.~\vsc\ in six stellar mass bins for galaxies with $0.005 < z < 0.075$. Points are color-coded according to \nuvr color as in Figure \ref{sd1_vdisp_plot}. In each panel, the solid line is a least-squares bisector fit to all galaxies in each mass bin. The slope of the best-fit relation is indicated. For reference, the dashed line is the global relation from Figure \ref{sd1_vdisp_plot}. The individual relations are consistent with that found by fitting to all galaxies simultaneously. The existence of a relation between \sphy\ and $\sigma_\mathrm{1}$ at fixed mass suggests that both quantities are indeed sensitive to mass increase in the \emph{inner} regions, i.e., to bulge buildup.}

	\label{sd1_vdisp_slice_plot}
\end{figure*}

To further investigate the relation between surface mass density and velocity dispersion, Figure \ref{sd1_vdisp_slice_plot} plots \sphy\ against \vsc\ in bins of stellar mass. In each panel, the best-fit least-squares line is included, and the slope of each relation is indicated. The slopes are all consistent with each other and with the value of 1.99 found in Figure \ref{sd1_vdisp_plot}. What this implies is that a tight correlation between inner surface density and velocity dispersion exists even at fixed stellar mass. In other words, both quantities are equally sensitive to central bulge buildup, and this behavior is independent of stellar mass. However, \sphy\ is attractive in a practical sense because it is more accurate even in lower-mass galaxies, where \vsc\ has large errors or cannot be measured reliably (in SDSS) at all. 

Recently, \citet{wake12} asserted that galaxy color is best correlated with velocity dispersion rather than with stellar mass, effective surface mass density, or Sersic index. Their conclusion was reached by examining the strength of \emph{residual} color correlations when one parameter is held fixed: only weak residual correlations between color and any of the remaining parameters were found at fixed velocity dispersion. But, such correlations manifestly do exist, especially for $\sigma\la 200\, \mathrm{km\,s^{-1}}$ (their Figure 4). To examine structural correlations with quenching, it is necessary to follow trends \emph{along evolutionary paths}, and thus at fixed stellar mass. Once this is done, \sphy\ and \vsc\ look equally predictive, except that \sphy\ is more accurately measured in small galaxies. 

\subsection{A New Black Hole Mass Scaling Relation?}

At low redshift, the new, tight relation between \sphy\ and \vsc\ might also be utilized in studies of the $M_\mathrm{BH}$-$\sigma$ relation. Again, since determining $\sigma$ is time-consuming, one might use \sphy\ as a proxy for velocity dispersion and reassess the implication for the properties of black holes in various galaxies. Alternatively, one might try to establish an entirely parallel $M_\mathrm{BH}$-$\Sigma_1$ relation and explore its implications. Calculating \sphy\ is made easy thanks to the wide availability of high-resolution images of nearby galaxies. Specifically, assuming a $M_\mathrm{BH}$-$\sigma$ relation of the form $M_\mathrm{BH}\propto\sigma^4$ \citep[e.g.,][]{tremaine02,gultekin09}, and substituting our relation \sphy$\,\propto\sigma_\mathrm{1}^{1.99}$, we find that $M_\mathrm{BH}\propto\Sigma_\mathrm{1}^{2.0}$. It will be interesting to verify if this prediction is indeed valid for nearby red sequence and green valley galaxies. It will also be interesting to see whether blue galaxies, which lie on the $\Sigma_1$-$\sigma_1$ relation, lie on or off the $M_\mathrm{BH}$-$\Sigma_1$ relation.
 
A $M_\mathrm{BH}$-$\Sigma_1$ relation could also be utilized to estimate the amount of growth in black hole mass as a galaxy evolves through the blue cloud and onto the red sequence. Such information would provide useful constraints on the coevolution of black holes and their host galaxies while they are still forming stars. This is particularly relevant given recent results indicating that AGNs may preferentially exist in star-forming galaxies \citep[e.g.,][]{mullaney12,trump13}. Assuming that galaxies evolve through the blue cloud at roughly constant stellar mass, Figure \ref{nuvr_sd1_plot} implies that a galaxy can increase its \sphy\ by a factor of $\sim4$ while evolving through the blue cloud. Using our predicted relation, $M_\mathrm{BH}\propto\Sigma_\mathrm{1}^{2.0}$, the black hole would then grow by a factor of $\sim16$.  
 
\section{DISCUSSION} \label{discussion}

Our discussion addresses several key questions raised earlier. We present alternative interpretations of the thresholds in structural parameters. A brief discussion about rejuvenated SF follows. Finally, we speculate on the possible mechanisms that quench SF and how they relate to the buildup of galaxy bulges and the properties of dark matter halos.

\subsection{Evolving Galaxies or Evolving Thresholds?}

Our interpretation of Figure \ref{nuvr_sd1_plot} is that galaxies increase their inner mass density (\sphy) during their evolution through the blue cloud. Once \sphy\ reaches a (mass-dependent) threshold value, galaxies are able to quench. An alternative interpretation is that a galaxy's \sphy\ is set at some early epoch and remains fixed through time and that the threshold value evolves down in time to ``sweep up'' galaxies, rather than galaxies building up \sphy\ over time. Indeed, \citet{franx08} and \citet{williams10} claim to find an evolving threshold surface density that decreases with time\footnote{More precisely, \citet{franx08} and \citet{williams10} use \emph{effective} surface density  ($\propto M_*/R_\mathrm{eff}^2$), and their samples are not divided into stellar mass bins when obtaining this result.}.

There are two arguments why the interpretation of an evolving threshold is less preferred. First, if the threshold were moving over time, one would expect to see signs of this in the \sphy\ distributions of more massive galaxies today. Assuming that the \sphy\ distributions of massive blue galaxies exhibited a spread comparable to that seen in present-day, \emph{less} massive blue galaxies (i.e., $\approx0.6$ dex total width; Figure \ref{nuvr_sd1_plot}), we would then expect to see a similarly broad \sphy\ distribution in their quenched descendants if \sphy\ is fixed as galaxies move through the green valley. This broad \sphy\ distribution in red sequence galaxies of higher mass is not seen. Rather, the total spread of red galaxies \emph{decreases} with mass from $\approx0.4$ dex in the lowest mass bin to $\approx0.3$ dex in the highest mass bin.

Another way of seeing this is that, if only the threshold moves and \sphy\ does not evolve, the motion of points in Figure \ref{nuvr_sd1_plot} would all be strictly vertical. Hence, the total width of the \sphy\ distribution from the bottom of the blue sequence to the top of the red sequence would remain the same in all mass bins. This is also not seen: in each mass bin, the width of the \sphy\ distribution narrows toward redder colors. Both viewpoints imply that the \sphy\ distribution either must start out narrower at higher mass, or, as is more plausible, galaxies are individually evolving upwards in \sphy\ and converging on the limiting value that is set by their stellar mass (Figure \ref{sd1_mass_plot}).

Second, galaxies do not live in a vacuum and are not frozen in time. Numerous processes exist which tend to make inner mass densities increase. These processes can be ranked according to whether they are rapid or slow, but all of them depend on the ability of dynamical subpopulations in the galaxy to exchange energy and angular momentum with one another. Examples from fast to slow include gas-rich major mergers \citep[e.g.,][]{Toomre72}, violent disk instabilities \citep[e.g.,][]{dekel09,ceverino10,cacciato12}, gas-rich minor mergers \citep[e.g.,][]{mihos94}, and bar and related non-axisymmetric disk instabilities \citep[e.g.,][]{kormendy04}. All of these processes raise inner mass density by driving gas toward the center in a dissipative manner. 

Thus, we \emph{know} that galaxy inner mass densities are increasing, the only question is by how much and how fast. To assess this, it is appropriate to note that our SDSS data reflect galaxies at the present epoch, which have mostly settled down into dynamical equilibrium. It is thus appropriate to consider bulge-building processes that are operative now, which leads us to focus on the slower processes listed above. Reliable rates for inner mass density buildup in mature disk galaxies require very accurate hydrodynamic simulations, which are only now becoming feasible. As an alternative, we cite the analytic toy model of \citet{forbes11}, which models the rather modest effects of gravitational disk instabilities. The net result from that process alone is a buildup in inner mass density by a factor $\sim2$ since $z\approx2$. This is consistent with the average difference in inner mass density between blue and red galaxies seen in our data (Figure \ref{sb_sd_prof_plot}), which suggests that this process could be a significant contributor. Also, \citet{patel13} present \emph{observational} evidence that blue, star-forming galaxies \emph{continually} grow their inner mass density as they evolve from $z\sim1$ to the present day. 

\subsection{The Utility of \sphy\ and \vsc\ as Quenching Predictors}\label{sect82}

The results presented above highlight the connection between inner mass density and the quenching of SF. It is not obvious whether the trends seen are causative or merely predictive of quenching. For now, we discuss the predictive nature of the correlation between color and \sphy\ (Figure \ref{nuvr_sd1_plot}). In this, we follow previous authors who also looked for structural parameters that are predictive of quenching \citep{kauffmann06,bell08,franx08,wake12,cheung12}. Those efforts were only partially successful because every parameter tried (e.g., stellar mass, effective surface density, or Sersic index) proved to show scatter, i.e., at a fixed value of the parameter, one could find both blue and red galaxies. In this respect, our new parameter \sphy\ has a similar weakness. Nevertheless, our approach has several merits worth discussing.  

A key feature of our analysis is the use of narrow mass slices to remove global trends with stellar mass. By doing so, we have been able to identify plausible evolutionary tracks (Figure \ref{nuvr_sd1_plot}). Such tracks would have been strongly blurred out if we had considered all galaxies together because of the strong trend between \sphy\ and stellar mass in Figure \ref{sd1_mass_plot}. In addition, our use of UV data from GALEX provides a clearer separation between star-forming and quenched galaxies. The improved dynamic range permits a more detailed study of galaxies that are \emph{currently} transitioning from blue to red, an advantage that has not been exploited previously.  

\begin{figure}
	\epsscale{1.2}
	\plotone{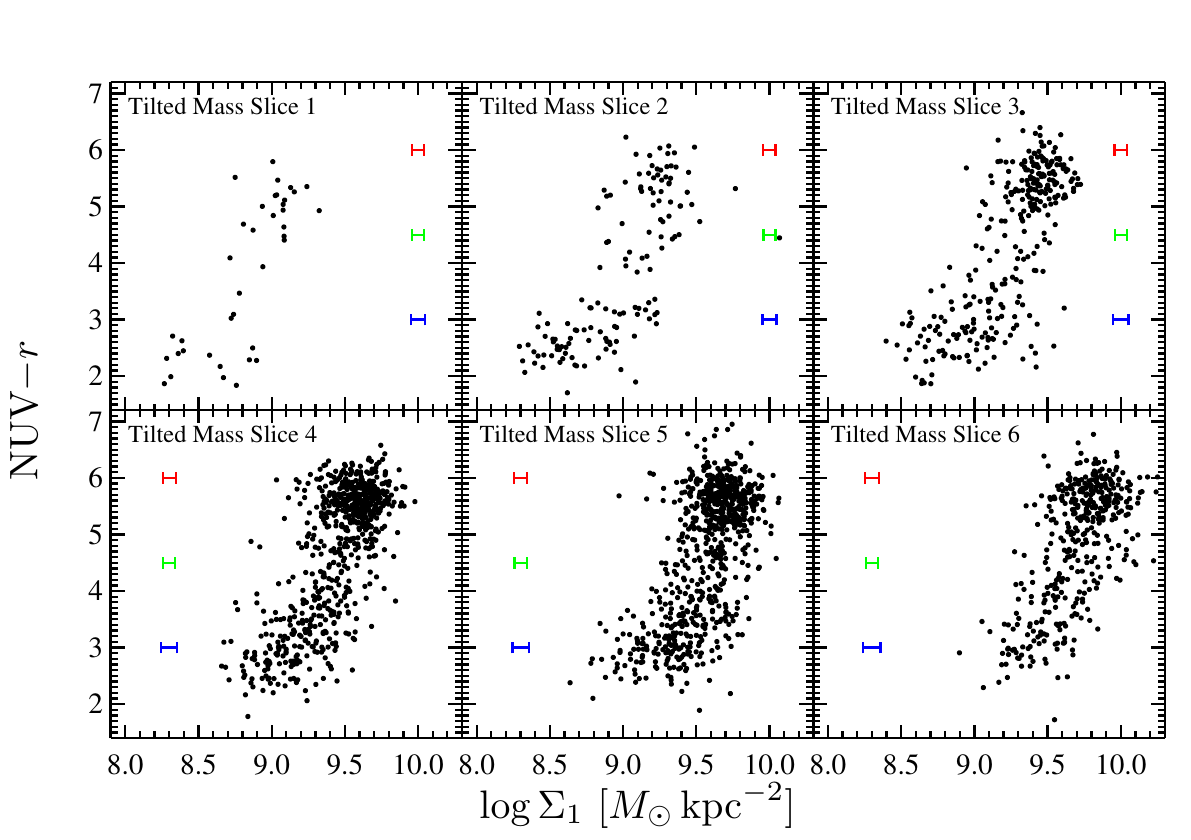}
	
	\caption{\nuvr vs. \sphy\ for the volume-limited sample. Each panel corresponds to a \emph{tilted} mass slice in the color-mass diagram, with the slope of the tilt corresponding to a growth in stellar mass by a factor of 2 from the blue sequence to the red sequence. Compared to Figure \ref{nuvr_sd1_plot}, the shape of the distribution in each panel remains nearly the same, implying that our results remain unchanged even when incorporating realistic mass growth rates explicitly. }

	\label{nuvr_sd1_tilt_plot}
\end{figure}

The validity of our assumption that galaxies quench at essentially fixed mass bears further discussion. In Section \ref{sd1_nuvr}, we appealed to simple physical arguments that show that galaxies do not increase their mass significantly over the expected quenching timescale ($\sim1$ Gyr). Indeed, the mass doubling time is several Gyr \citep[from $z\approx1$ to 0;][]{marchesini09,behroozi13}, suggesting that our assumption of evolution through the green valley at constant mass is reasonable. Nevertheless, it is worth verifying explicitly how such mass growth would affect our results. Figure \ref{nuvr_sd1_tilt_plot} presents the relation between \nuvr and \sphy\ for galaxies in our sample, this time divided into mass bins that are tilted in the color-mass diagram, with the slope of the tilt corresponding to a growth in stellar mass by a factor of 2 from the blue cloud to the red sequence. Compared with Figure \ref{nuvr_sd1_plot}, it is apparent that the overall distributions remain essentially unchanged, and the total increase in \sphy\ along an evolutionary path is, if anything, increased. Thus, our results are robust even when incorporating realistic mass growth rates explicitly.

Other potentially valid structural parameters include total stellar mass and inner velocity dispersion. However, the former is known to be a poor predictor of quenching \citep{franx08,bell12,cheung12}, and the latter is seemingly equivalent to \sphy\ in its predictive power (see below).

Thus, the situation is fundamentally different for color overlap in \sphy\ compared to color overlap in stellar mass. The latter reflects the fact that mass by itself is a genuinely poor predictor of color whereas the former (at fixed stellar mass) reflects the fact that galaxies evolve rapidly in color after attaining a certain value of \sphy. The key difference is that galaxies evolve in \sphy\ (and color changes) in a single mass bin whereas the locus of points in the color-mass diagram does \emph{not} reflect a \emph{single} evolutionary track.  

A benefit of using \sphy\ as a tracer of quenching is that it is measured from galaxy \emph{mass} profiles, rather than from \emph{light} profiles. As shown in Figures \ref{sb_sd_prof_plot} and \ref{radii_ratio_plot}, the surface brightness profiles exaggerate the difference in structure between blue and red galaxies. This impacts measurements of other structural parameters (e.g., effective surface density, effective velocity dispersion, and Sersic index), which are traditionally based on light profiles. In other words, this ``outshining'' may exaggerate previously seen trends between color and these parameters, such that the strong differences between the structure of red and blue galaxies is simply dependent on whether a galaxy is star-forming or not. Using mass-weighted versions of these parameters would decrease the apparent strength of these trends. With the availability of high-resolution data at increasingly large redshifts, the use of mass profiles in characterizing structure is becoming more common in galaxy evolution studies \citep[e.g.,][]{vdk10,wuyts12,patel13,szomoru13}.

\subsection{On the Verticality and Scatter of Evolutionary Tracks}

As seen in Figures \ref{nuvr_sd1_plot} and \ref{nuvr_vdisp_slice_plot}, green and red galaxies trace out a nearly vertical distribution in each mass bin. What are the implications of the verticality of these plausible evolutionary tracks? What does the width of the distribution, particularly in the green valley, tell us about the relation between galaxy structure and quenching? 

To address these questions, suppose that the distribution in Figure \ref{nuvr_sd1_plot} were a \emph{perfectly narrow} step function, but that \sphy\ has scatter caused by measurement errors. The step function would then be blurred, and both red and blue galaxies would be found at the same value of \sphy. Thus, under a highly non-linear trend like a step function, standard measures of prediction accuracy, such as the relative number of red and blue galaxies at fixed \sphy, would tend to break down due to the confounding effects of measurement error on an intrinsically narrow distribution.

An alternative way to view this is relative timescales. If galaxies quench at fixed mass, the verticality of the tracks in Figure \ref{nuvr_sd1_plot} is due to rapid evolution in color after reaching a critical value of \sphy. Thus, it is impossible, in principle, to find a structural parameter that perfectly predicts the quenching state (color) if the quenching timescale is very rapid compared to the evolution timescale of the relevant structural parameter. In other words, the change in structure ``lags behind'' the change in galaxy color.

One way to proceed is to ask whether it is possible to narrow the scatter still further below that seen in Figure \ref{nuvr_sd1_plot}. If so, this may indicate that galaxies indeed evolve over a step function in \sphy\ and that the observed scatter is (solely) due to measurement error. We attempted to reduce the scatter by retreating more closely to the original data. Specifically, instead of \sphy, we used the average $i$-band surface brightness, $\mu_1$, within a 1-kpc radius aperture. This eliminates the $M/L$ conversion, which has a scatter of 0.04 dex. To reduce other sources of uncertainty, we selected a narrowly defined subsample for this test: only galaxies in our volume-limited sample with $10.25<\log\, M_*/M_\odot<10.5$ located between $0.045<z<0.055$ were chosen. The chosen redshift range permits us to calculate $\mu_1$ without having to interpolate between SDSS apertures. The scatter in $\mu_1$ for the green valley is 0.151 dex ($0.38\, \mathrm{mag\,arcsec}^{-2}$), comparable to the 0.154 dex seen in Figure \ref{nuvr_sd1_plot} for galaxies in the same mass bin. Thus, it appears that the scatter in \sphy\ is not due solely to these measurement uncertainties. 

How does velocity dispersion fare in this regard? As seen in Figure \ref{nuvr_vdisp_slice_plot}, the typical rms scatter in \vsc\ for green valley galaxies is $\la0.1$ dex. In order to make a fair comparison with \sphy, the scatter must be multiplied by 2 (the slope of the \sphy-\vsc\ relation in Figure \ref{sd1_vdisp_plot}). Doing so, we find that the scatter in \vsc\ is larger than that for \sphy\ for lower-mass galaxies ($<10^{10.5} M_\odot$), but is comparable to or smaller than that for \sphy\ for high-mass galaxies ($>10^{10.5} M_\odot$). The reduced scatter in \vsc\ would suggest that \vsc\ offers a slightly cleaner measure of the intrinsic scatter in the green valley, at least for high-mass galaxies.  

It thus appears that the intrinsic scatter in \sphy\ and \vsc\ through the green valley is not zero, and hence it is unclear if a narrow step function is a correct description of the trends seen in Figures \ref{nuvr_sd1_plot} and \ref{nuvr_vdisp_slice_plot}. To be sure, we have not accounted for all possible factors that could contribute to the scatter (e.g., dust effects, inclination, rejuvenated SF [see below]). It would be interesting to investigate systematically these effects with, e.g., semi-analytic models to better understand the behavior observed in Figures \ref{nuvr_sd1_plot} and \ref{nuvr_vdisp_slice_plot}. Regardless, our results present useful constraints that theoretical models of galaxy evolution need to match.

\subsection{Constraints on Rejuvenated SF}

While the dominant ``flow'' of galaxies is believed to be from blue to red \citep[e.g.,][]{bell04,faber07,martin07}, an increasing number of observations have uncovered evidence of previously quiescent galaxies undergoing ``rejuvenated'' SF as a result of recently acquired gas \citep[e.g.,][]{kaviraj07,salim10,thilker10,lemonias11,marino11,salim12,fang12}. Such a rejuvenated galaxy is driven from the red sequence back into the green valley (or blue sequence). \citet{kauffmann06} hypothesized that galaxies with effective surface densities below the threshold value of $\sim10^{8.5}\, M_\odot\, \mathrm{kpc}^{-2}$ undergo increasingly bursty episodes of SF triggered by stochastic gas accretion that can drive red galaxies toward bluer colors. Moreover, these rejuvenation events were predicted to be so rapid and frequent that the color bimodality of galaxies at fixed mass (or effective surface density) was due \emph{purely} to jumping between the blue and red states. 

While our results do not rule out that rejuvenation is occurring in some galaxies, we can constrain how it must proceed. We assume first that the stellar mass growth during rejuvenation is small so that the constant-mass slices in Figure \ref{nuvr_sd1_plot} still apply. Figure \ref{nuvr_sd1_plot} then shows that galaxies can move from red to blue (at fixed mass) as long as no significant central mass growth occurs. In other words, rejuvenated galaxies are allowed to move vertically in Figure \ref{nuvr_sd1_plot}, with the consequence that any new SF occurs mainly in the outer parts. This is indeed seen in UV images of candidate rejuvenated galaxies, where young stars are seen to form in extended disks and rings \citep {kauffmann07,salim10,thilker10,salim12,fang12}. Extended UV emission is also seen in some low surface brightness galaxies, and their UV colors are also consistent with rejuvenated SF \citep{boissier08}.

According to this interpretation, at least some of the vertical spread in \nuvr color at fixed \sphy\ in Figure \ref{nuvr_sd1_plot} may be caused by rejuvenation. This is an interesting hypothesis as it picks out certain galaxies that are prime candidates for rejuvenation, namely, those with high \sphy\ but blue colors, and these could be studied further. However, an equally important use of Figure \ref{nuvr_sd1_plot} is to \emph{limit} the amount of transiting back and forth between the blue and red sequences that can occur. In particular, increasing \sphy\ is a one-way trip---it cannot be undone, at least not easily. Hence, the \emph{net} difference in \sphy\ between blue and red galaxies is important; if all galaxies were making multiple trips back and forth between the red and blue sequences, we would expect their \sphy\ to be more equal, and this is not seen. Even stronger constraints may come from Figure \ref{nuvr_vdisp_slice_plot}, which uses velocity dispersion rather than \sphy. The tilt of the ``hooks'' is larger in this figure, further limiting the amount of hopping back and forth that can be taking place at (presumably) fixed $\sigma$. The overall conclusion is that the new data permit some degree of rejuvenation, but it cannot be a wholesale phenomenon. Better models and better understanding of the horizontal scatter in Figures \ref{nuvr_sd1_plot} and \ref{nuvr_vdisp_slice_plot} may be able to tightly limit this process in the future.

\subsection{Is Quenching a Two-Step Process?}

As discussed in the introduction, several processes have been proposed that can shut down SF in galaxies by stabilizing, expelling, and/or heating up any gas in the galaxy or surrounding halo. Broadly speaking, these mechanisms can be divided into two classes: those that are governed by the conditions of the dark matter halo (virial shock-heating of gas) and those that are driven by bulge buildup (e.g., morphological quenching, AGN feedback, and stellar and supernova feedback in a central starburst). Conventional wisdom typically assumes that any one of these processes working alone is sufficient to quench a galaxy's SF. However, the results presented in this paper necessitate a closer examination of this assumption. 

We have emphasized that our results imply that galaxy quenching requires the presence of a bulge, but that a bulge alone is not enough to ensure the complete cessation of SF. This is reflected in e.g., the large vertical scatter in color in Figure \ref{nuvr_sd1_plot}: galaxies with high \sphy\ can be both blue or red. In addition, the critical value of \sphy\ above which galaxies can quench is seen to be a strong function of stellar mass (Figure \ref{sd1_mass_plot}). Both observations suggest that a single structural parameter is insufficient to fully predict the quenching state of a galaxy; two parameters may be necessary. 

But if one parameter is indeed insufficient, what would be the physical basis for two? A possible answer may be found by realizing that genuine quenching requires that \emph{two} conditions be met: 1) gas internal to a galaxy must be heated/expelled \emph{and} 2) external gas accretion must be permanently halted. The first requirement that internal gas be prevented from forming stars can be achieved by the influence of a strong bulge and its attendant AGN feedback and/or disk stabilization \citep[e.g.,][]{dimatteo05,croton06,martig09}, or supernova feedback from a central starburst triggered by inflowing gas \citep[e.g.,][]{dekel86,hopkins06}. The state of the bulge is thus one physical property that is plausibly linked to quenching, and can be traced by a parameter like \sphy. Preventing subsequent gas accretion onto the galaxy, however, is a halo-regulated process; gas becomes shock-heated once the halo reaches some critical mass \citep[$\sim10^{12}\,M_\odot$;][]{birnboim03,dekel06}. Thus, our sought-after second parameter must be able to trace the state of a galaxy's dark matter halo. Direct observational measurements of halo properties (mass, size, etc.) are impossible, and hence indirect indicators are needed. Candidate parameters include a galaxy's velocity dispersion \citep{wake12a} or its total stellar mass \citep{li13}. 

Having recognized that quenching is a two-step process best described using two structural parameters, the next issue is to understand the connection between bulge-driven and halo-driven quenching processes. In other words, how does the bulge know about the halo, or vice versa? We speculate on some potential methods to link the two together. Recent galaxy simulations have shown that the formation of a stable virial shock can be triggered by a minor merger that sets off a shock wave in the unstable gas in a halo around the critical mass (A.~Dekel, private communication). In addition, the minor merger can bring gas into the center of the galaxy, form additional stars (increasing \sphy), and perhaps trigger an AGN. Another possibility is that AGN feedback is most effective at heating and driving out gas only in halos that have reached the critical mass \citep{dekel06}.

To summarize, the results of this paper and related work point to the importance of the bulge in quenching, namely, that the growth of the bulge is a necessary condition for a galaxy to be quenched, whether by stabilizing or removing any gas. However, in order to prevent further gas accretion, the halo must also be massive enough to maintain a shock that heats infalling gas and prevents it from cooling onto the galaxy. Moreover, if these two processes do not happen in perfect synchrony, it could result in the horizontal scatter seen in e.g., Figure \ref{nuvr_sd1_plot}. A clear quenching threshold would only emerge in a two-dimensional parameter space that \emph{combines} the operative bulge and halo properties. We have not yet explored what this two-dimensional parameter space looks like. The present paper is a first step, but our measures of halo properties, such as stellar mass or velocity dispersion, are still rather indirect. The next step is to extend the present analysis using a more direct measure of halo properties, such as halo mass, to see if the predictions sharpen further. Such an analysis is in progress (J.~Woo et al.~2013, in preparation). 

\section{SUMMARY AND CONCLUSIONS} \label{conclusions}

This paper studies the connection between the quenching of star formation and the growth of the stellar mass surface density in bulges of a sample of central SDSS galaxies with $9.75<\log M_*/M_\odot<11.25$ and $z<0.075$. Bulge growth is traced by the stellar mass surface density within a radius of 1 kpc, \sphy, and star formation is traced through \nuvr color, which can select galaxies in the green valley that are \emph{currently} transitioning from blue to red. A key and novel aspect of our work is dividing the sample into narrow mass slices, which plausibly selects groups of galaxies that are traveling along nearly the same evolutionary track. Our main results are as follows.

\begin{enumerate}

\item The critical value of \sphy\ above which galaxies are predominantly quenched increases with stellar mass, $\Sigma_1\propto M_*^{0.64}$; it is not a fixed universal value. This suggests that at least two structural parameters that are not fully correlated are necessary to shut down star formation.

\item At fixed stellar mass, a galaxy's color is closely related to \sphy. Specifically, \sphy\ seems to increase in the blue cloud until it reaches a mass-dependent threshold value, then star formation quenches, and color starts to redden. Reaching this threshold is a necessary, but not sufficient, condition to be quenched. 

\item The surface brightness profiles of blue galaxies are brighter (fainter) in the outer (inner) parts compared to green and red galaxies of the same stellar mass. However, the outer \emph{mass} density profiles are remarkably similar for all galaxies in a given mass slice regardless of color. The one difference is a slightly lower mass surface density in blue galaxies by a factor of $\sim2$--3 within 1 kpc. This indicates that the growth of the bulge is the key structural change related to galaxy quenching. In addition, the common use of light profiles to measure quantities like sizes exaggerate structural differences between blue and red galaxies; stellar mass profiles should be used instead.

\item A strong correlation between \sphy\ and velocity dispersion \vsc\ is seen for galaxies of all colors, $\Sigma_1\propto \sigma_1^{2.0}$. The scaling is consistent with simple dynamical arguments and highlights the fact that both parameters are useful tracers of the central bulge. Using this result, we predict a black hole mass scaling relation of the form $M_\mathrm{BH}\propto\Sigma_1^{2.0}$, in analogy to the established $M_\mathrm{BH}\propto\sigma^4$ relation. 

\item Our finding that at least two structural parameters are needed to predict quenching (e.g., stellar mass and \sphy\ or \vsc) may be able to reconcile the two broad classes of quenching theories. On the one hand, gas already inside the galaxy must be expelled, heated, or stabilized from forming stars. This process (e.g., AGN feedback, morphological quenching) requires a sufficiently dense bulge and can be traced by \sphy. This by itself is not sufficient to guarantee quenching, however. Gas accretion onto the galaxy must also be halted, and the relevant process for this (virial shock-heating) activates once the halo reaches a threshold mass. 

\end{enumerate}

The results of this work highlight the need to better understand the interplay between the inner conditions of galaxies and their surrounding dark matter halos during the quenching process. In addition, the scaling relations presented here offer useful constraints on the structural evolution of galaxies that can be compared to theoretical models. It will also be interesting to study the behavior of high-redshift galaxies in these parameter spaces and examine how they differ from the galaxies studied here.
 
\acknowledgements

We gratefully thank the anonymous referee for prompt and helpful comments that improved the paper. We thank Eric Bell, Edmond Cheung, Charlie Conroy, Joanna Woo, and Hassen Yesuf for useful conversations regarding this work. We acknowledge financial support from NSF grant AST-0808133. A.~D.~also acknowledges support from ISF grant 24/12, GIF grant G-1052-104.7/2009, a DIP grant, and NSF grant AST-1010033.

Funding for the SDSS and SDSS-II has been provided by the Alfred P. Sloan Foundation, the Participating Institutions, the National Science Foundation, the U.S. Department of Energy, the National Aeronautics and Space Administration, the Japanese Monbukagakusho, the Max Planck Society, and the Higher Education Funding Council for England. The SDSS Web Site is http://www.sdss.org/.

GALEX (Galaxy Evolution Explorer) is a NASA Small Explorer, launched in 2003 April. We gratefully acknowledge NASA's support for construction, operation, and science analysis for the GALEX mission, developed in cooperation with the Centre National d'\'Etudes Spatiales of France and the Korean Ministry of Science and Technology.

{\it Facilities:} \facility{GALEX}, \facility{Sloan}



\appendix

\section{SEEING EFFECTS ON \sphy\ MEASUREMENTS} \label{redshift_bias}

\begin{figure}[h]
	\epsscale{0.8}
	\plotone{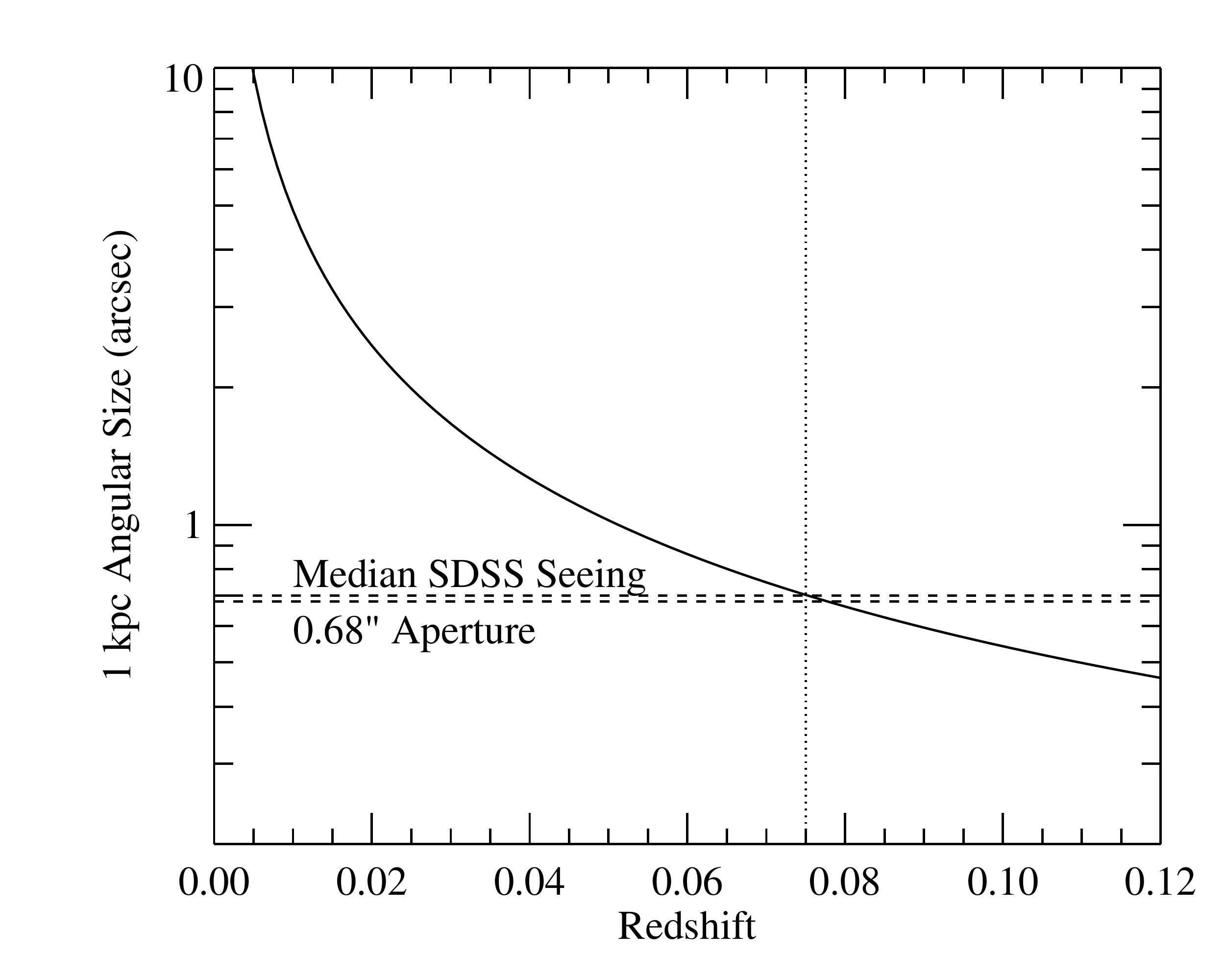}
	
	\caption{The angular size (in arcsec) corresponding to 1 kpc physical size as a function of redshift. The upper dashed line indicates the median SDSS seeing (HWHM = 0\farcs7). The lower dashed line corresponds to the smallest aperture ($R=0\farcs68$) used in the SDSS pipeline that is comparable in size to the seeing. To reduce scatter in our measurements of \sphy\ due to seeing, we restrict the redshift range to $z<0.075$ (vertical dotted line) for the most massive galaxies when defining the volume-limited sample.}

	\label{aper_red_plot}
\end{figure}

A concern is that the resolution of the SDSS imaging may compromise the reliability of our measurements of \sphy, especially at the highest redshifts considered in the volume-limited sample (Figure \ref{complete_plot}). Here we motivate our adoption of $z=0.075$ as the maximum allowed redshift for the two highest mass bins. Figure \ref{aper_red_plot} shows the angular extent corresponding to 1~kpc physical size as a function of redshift. Two dashed lines are plotted indicating (1) the SDSS seeing HWHM of 0\farcs7 and (2) the smallest aperture radius used in the SDSS pipeline ($R=0\farcs68$) that is comparable to the seeing. As can be seen, the 0\farcs68 aperture photometry is usable out to $z=0.075$, and this value is adopted as the maximum redshift limit to ensure reliable values of \sphy\ for the most massive galaxies in the sample.

\begin{figure}
	\epsscale{1}
	\plotone{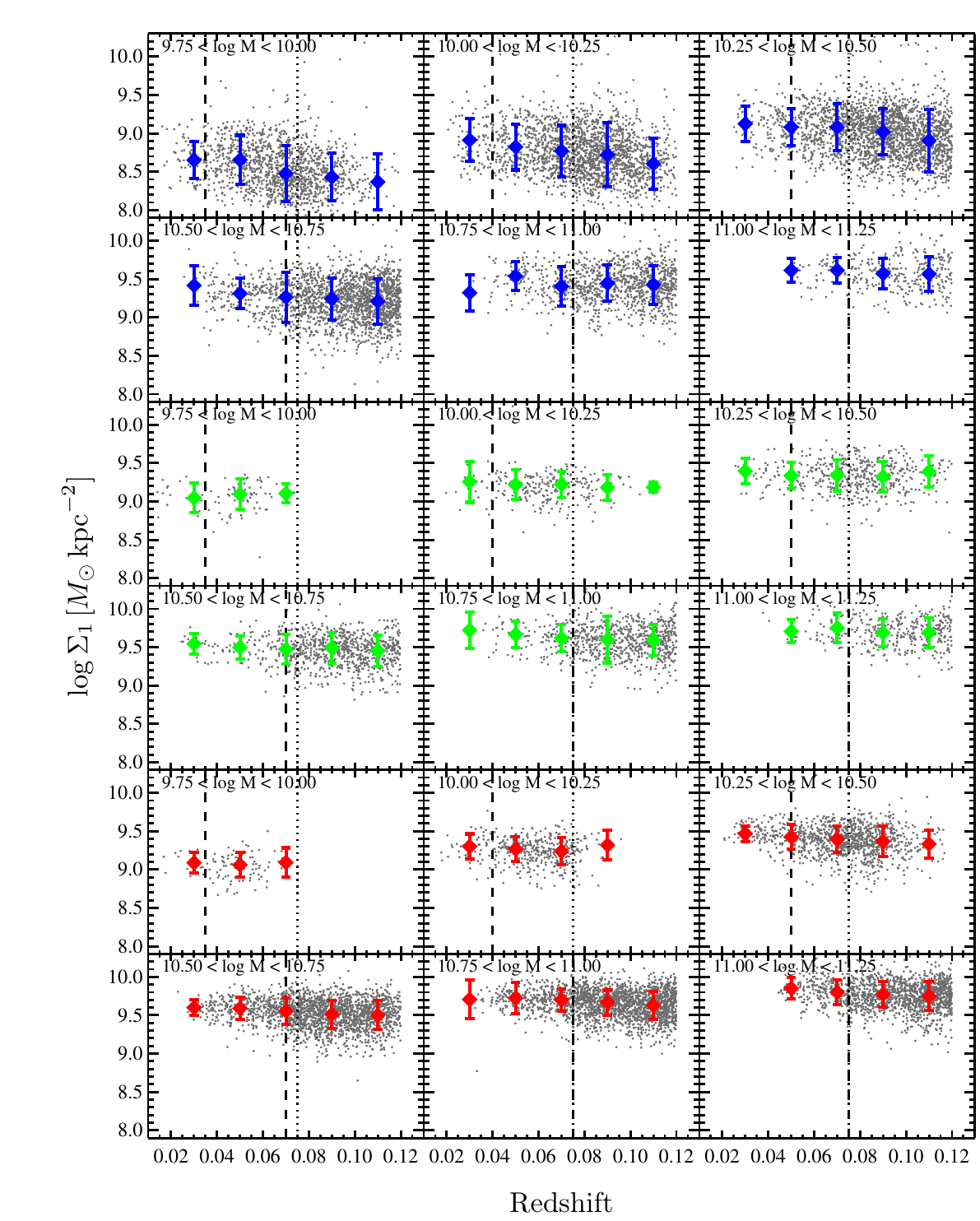}
	
	\caption{\sphy\ vs.~redshift for blue (\emph{top}), green (\emph{middle}), and red (\emph{bottom}) galaxies in six stellar mass bins. Diamonds indicate the median \sphy\ in five redshift bins, and the error bars represent the $1\sigma$ scatter. The vertical dashed line in each panel indicates the maximum redshift used to define the volume-limited sample for each mass bin (Figure \ref{complete_plot}). The median and scatter in \sphy\ remain essentially constant within the redshift limits for each mass bin. The dotted line indicates the redshift ($z=0.075$) where the size of the PSF is comparable to 1 kpc.}

	\label{redshift1_plot}
\end{figure}

To check the validity of this redshift limit, Figure \ref{redshift1_plot} presents the values of \sphy\ as a function of redshift for blue, green, and red galaxies. The diamonds indicate the median value of \sphy\ in redshift bins, and the error bars indicate the $1\sigma$ scatter of \sphy\ in each bin. In each panel, the vertical dashed line indicates the redshift limit adopted for each mass bin in defining the volume-limited sample (Figure \ref{complete_plot}). The vertical dotted line indicates the redshift $z=0.075$, above which seeing affects the reliability of measurements of \sphy. For galaxies of all colors, the median value of \sphy\ and the scatter are essentially constant within the redshift ranges adopted, and even out to $z=0.075$.

\end{document}